\documentclass[useAMS,usenatbib]{mn2e}
\usepackage{graphicx}
\usepackage{amsmath}
\usepackage{bm}
\usepackage{multirow}

\def\ltsima{$\; \buildrel < \over \sim \;$}
\def\ltsim{\lower.5ex\hbox{\ltsima}}
\def\gtsima{$\; \buildrel > \over \sim \;$}
\def\gtsim{\lower.5ex\hbox{\gtsima}}

\title[]{Impact on the tensor-to-scalar ratio of incorrect Galactic foreground modelling}
\author[C. Armitage-Caplan et al.]
{Charmaine Armitage-Caplan,$^1$\thanks{armitage-caplan@physics.ox.ac.uk}
Joanna Dunkley,$^1$
Hans Kristian Eriksen,$^2$
\newauthor
Clive Dickinson,$^3$\\
$^1$Sub-department of Astrophysics, University of Oxford, Denys Wilkinson Building, Keble Road, Oxford OX1 3RH, UK\\
$^2$Institute of Theoretical Astrophysics, University of Oslo, P.O. Box 1029 Blindern, N-0315 Oslo, Norway\\
$^3$Jodrell Bank Centre for Astrophysics, Alan Turing Building, School of Physics \& Astronomy, \\The University of Manchester, Oxford Road, Manchester M13 9PL, UK
}

\date{}

\begin{document}

\label{firstpage}

\maketitle

\begin{abstract}
A key goal of many Cosmic Microwave Background experiments is the detection of gravitational waves, through their B-mode polarization signal at large scales. To extract such a signal requires modelling contamination from the Galaxy. Using the {\it Planck} experiment as an example, we investigate the impact of incorrectly modelling foregrounds on estimates of the polarized CMB, quantified by the bias in tensor-to-scalar ratio $r$, and optical depth $\tau$.
We use a Bayesian parameter estimation method to estimate the CMB, synchrotron, and thermal dust components from simulated observations spanning 30-353~GHz, starting from a model that fits the simulated data, returning $r<0.03$ at 95\% confidence for an $r=0$ model, and $r=0.09\pm0.03$ for an $r=0.1$ model. We then introduce a set of mismatches between the simulated data and assumed model. Including a curvature of the synchrotron spectral index with frequency, but assuming a power-law model, can bias $r$ high by $\sim1\sigma$ ($\delta r\sim0.03$). A similar bias is seen for thermal dust with a modified black-body frequency dependence, incorrectly modelled as a power-law. If too much freedom is allowed in the model, for example fitting for spectral indices in 3 degree pixels over the sky with physically reasonable priors, we find $r$ can be biased up to $\sim 3\sigma$ high by effectively setting the indices to the wrong values. Increasing the signal-to-noise ratio by reducing  parameters, or adding additional foreground data, reduces the bias. We also find that neglecting a $\sim1\%$ polarized free-free or spinning dust component has a negligible effect on $r$. These tests highlight the importance of modelling the foregrounds in a way that allows for sufficient complexity, while minimizing the number of free parameters.

\end{abstract}
\nokeywords
\section{Introduction}
Extraction of the polarized Cosmic Microwave Background signal at large scales is hampered by significant levels of polarized Galactic emission. The two dominant components are synchrotron and thermal dust, polarized due to the coherent magnetic field in the Galaxy \citep[e.g.,][]{Page07,Fraisse08}. For an experiment observing at multiple frequencies, one method of separating the signals is to parameterize the synchrotron and dust, and to fit for these components, in addition to the CMB, over the region of sky where Galactic emission is lowest. While demonstrated to work for E-mode polarization \citep{Page07,Dunkley-WMAP,Gold09}, the signal of interest is the much smaller B-mode signal from inflation \citep[e.g.,][]{basko/polnarev:1980,bond/efstathiou:1984}. A concern with using such methods is that an incorrect model can lead to bias in the estimated CMB signal.

The {\it Planck} satellite mission, launched in May 2009, is measuring the polarization signal of the CMB in seven channels over the frequency range 30-353 GHz \citep{Planck06,Tauber2010,PlanckMission:2011}. While {\it Planck} will produce polarization
data, which offer a multitude of opportunities including possible recovery of inflationary B-modes at large scales and greater understanding of the polarized nature of Galactic foregrounds, it also comes with great challenges.  For an all-sky experiment like {\it Planck}, component separation of the polarization signal is more difficult than for the temperature counterpart, in part because the ratio of the foreground signal to CMB signal is higher. 

In many simulated tests of component separation, the simulations of the Galactic emission are well matched to the model used to describe them. Using a Bayesian component separation method which allows us to assume different models of the Galactic signal, we explore the effect on the recovery of the CMB in varying scenarios of mismatch between the model and simulation.  We use the recovered CMB map and its covariance to estimate two cosmological parameters: the optical depth to reionization, $\tau$, and the tensor-to-scalar ratio, $r$.  In this way, we can directly quantify the bias generated in the parameter estimation as a result of any particular model-simulation mismatch. Both the Bayesian component estimation method and the simulated skies used in this paper were first used and described in a previous paper, \citet{ArmitageCaplan11}.  There we examined the prospects for large-scale
polarized map and cosmological parameter estimation with simulated {\it Planck}
data for a single model-simulation combination.  This paper is a natural extension in which we use the same methods to recover maps and estimate parameters, while varying the simulated data and separation model.

In \S\ref{sec:methods}, we provide a brief overview of the Gibbs sampling method, the subsequent processing of the sampled distribution, and the likelihood estimation method.  In \S\ref{sec:psm}, we explain how the data are simulated.  A detailed account of the mismatch tests that we examine, and their resulting parameter estimates, is then presented in \S\ref{sec:mismatch}.  We then discuss the results and methods for mitigating possible biases in \S\ref{sec:discuss}, and conclude in S\ref{sec:conclude}.

\section{Method}
\label{sec:methods}

In the Bayesian parameter estimation method of foreground removal, the emission models of the CMB and foregrounds are parametrized based on our understanding of their frequency dependence. Focusing on polarization analysis, a sampling method is then used to estimate the marginalized CMB Q and U Stokes vector maps (and additionally the marginalized foreground maps) in every pixel over the sky. In general this extends template-removal methods to allow for spatial variation of the foreground spectral indices, and was first used to clean WMAP polarization data in \citet{Dunkley-WMAP}.  In this analysis, we use HEALPix \citep{Healpix} $N_{\rm side} = 16$ maps containing $N_p = 3072$ pixels.    
We use a code called {\it Commander} (see \citet{E06} and \citet{E08}) to perform the Gibbs sampling. 
The sampled distribution is then processed into a mean map and covariance matrix.  Finally, we perform a likelihood estimation for the two cosmological parameters, $\tau$ and $r$.
 
\subsection{Bayesian Estimation of sky maps}
\label{sec:bayes_est}

By Bayes' theorem, the posterior distribution for parameters, $\mathbf{s}$, given a set of maps, $\mathbf{d}$, can be written as
\begin{equation}
P(\mathbf{s}|\mathbf{d}) \propto P(\mathbf{d}|\mathbf{s}) P(\mathbf{s})
\end{equation}
with a prior distribution for the model parameters, $P(\mathbf{s})$. The Gaussian likelihood of the observed maps is given by
\begin{equation}
-2 \mathrm{ln} P(\mathbf{d}|\mathbf{s}) = \sum_{\nu}[\mathbf{d}_{\nu}-\mathbf{s}_{\nu}]^T\mathbf{N}_{\nu}^{-1}[\mathbf{d}_{\nu}-\mathbf{s}_{\nu}]
\end{equation}
where $\mathbf{d}_{\nu}$ is the observed sky map at frequency $\nu$, and $\mathbf{N}_{\nu}$ is its covariance matrix.

As in \citet{Dunkley-WMAP,ArmitageCaplan11}, we assume that the polarized Galactic emission is dominated by synchrotron and dust emission, arising due to the orientation of the Galactic magnetic field \citep[e.g.,][]{Page07}. We define a parametric model for the total sky signal in antenna temperature for a three-component model ($k=1$ for CMB, $k=2$ for synchrotron emission, and
 $k=3$ for thermal dust emission) as
\begin{equation}
\label{eq:model}
\mathbf{s}_{\nu} = \bm\alpha_1(\nu)\mathbf{A}_1 + \bm\alpha_2(\nu;\beta_2)\mathbf{A}_2 + \bm\alpha_3(\nu;\beta_3)\mathbf{A}_3
\end{equation}where $\bm{A}_k$ are amplitude vectors of length $2N_p$ and $\bm\alpha_k(\nu;\beta_k)$ are diagonal coefficient matrices of side $2N_p$ at each frequency. 

Once our model, and priors on the model parameters, are defined, we estimate the joint CMB-foreground posterior $P(\mathbf{A},\bm{\beta}|\mathbf{d})$ from which we can then obtain the marginalized distribution for the CMB map vector,
\begin{equation}
p(\mathbf{A}_{1},\mathbf{d}) = \int p(\mathbf{A},\bm\beta|\mathbf{d})d\mathbf{A}_2 d\mathbf{A}_3 d\bm{\beta}
\end{equation}
and similarly for the other model parameters.  

For the multivariate problem that we are considering, Gibbs sampling draws from the joint distribution by sampling each parameter conditionally as follows
\begin{align}
\mathbf{A}^{i+1} & \leftarrow P(\mathbf{A}|\bm\beta,\mathbf{d})\\
\label{eq:beta_samp}
\bm\beta^{i+1} & \leftarrow P(\bm\beta|\mathbf{A},\mathbf{d}).
\end{align}
We use {\it Commander} to implement the sampling of the amplitude-type and spectral index parameters.
{\it Commander} is a flexible code for joint component separation and CMB power spectrum
estimation; the reader is directed to \citet{ArmitageCaplan11} for a full description of its use for sampling only the sky signal.

\subsection{Likelihood estimation of cosmological parameters}

The product of a Bayesian parametric map estimation method is both a CMB map (which is taken to be the mean map calculated from the Gibbs chain after some burn-in) and a covariance matrix (which can be estimated from the marginalized posterior 
distribution) and together these products can be used to place constraints on cosmological parameters.  We compute the likelihood of the estimated maps, given a theoretical angular power spectrum, using the exact pixel-likelihood method described in \citet{Page07,ArmitageCaplan11}.  

The two cosmological parameters constrained by the large scale CMB polarization signal are the optical depth to reionization, $\tau$, and the 
tensor-to-scalar ratio, $r$.  The signature of reionization is at $\ell \ltsim 20$ in $C_{\ell}^{EE}$ where the amplitude of the reionization signal is proportional to $\tau^2$.  The tensor-to-scalar ratio $r$ directly scales the $C_{\ell}^{BB}$ power spectrum and is best probed at two angular scales: at the low $\ell\ltsim 20$ `reionization bump' before $C_{\ell}^{BB}$ due to lensing dominates, or at the smaller scale $\ell \sim 100$ `recombination bump' where foregrounds are expected to be lower but lensing is a contaminant. In this study we are considering constraints from the large-scale reionization bump, using $\sim$75\% of the sky.

By varying only the optical depth to reionization, and fixing the temperature anisotropy power at the first acoustic peak ($\ell=220$), we calculate the likelihood for each value of $\tau$. Separately, we vary only the tensor-to-scalar ratio, and calculate the likelihood at each value of $r$. The resulting one-dimensional distributions for $r$ and $\tau$ then include marginalization over foreground uncertainty.  To account for imperfect foreground cleaning in the Galactic plane, we apply a Galactic mask when calculating the likelihoods. In this analysis we use the standard WMAP `P06' mask \citep{Page07}, which masks $26\%$ of the sky.

\section{Simulated maps}
\label{sec:psm}

We generate simulated maps at the seven polarized nominal frequency channels for Planck (30, 44, 70, 100, 143, 217, and 353 GHz).  In our analysis, we do not apply beams or smoothing to the data; these would be included in a more realistic analysis but are not expected to significantly affect results.  
Realizations of the CMB are generated from a power spectrum computed using $\Lambda$CDM cosmological parameters \citep{Komatsu2011}, with either $r=0$ or $r=0.1$. 
Diagonal white noise realizations are generated based on the noise levels taken from the {\it Planck} Bluebook \citep{Planck06}, and we scale the given noise levels at beam-sized pixels to the corresponding noise level at $N_{\rm side}=16$ sized pixels, with side $3.6^\circ$.  This noise model is over-simplified as it contains no $1/f$-noise or other spatial correlations that are reported in the `early' Planck papers, which would increase effective noise levels \citep{PlanckHFI:2011,Zacchei2011}.

For the foreground components, we use two baseline tests to benchmark the level of bias in the mismatch tests.

In Test 1 (baseline with uniform $\beta_s$), spectral indices given by simple power-laws are used to simulate the synchrotron and dust foregrounds, and as a model in the component estimation.  
The simulated synchrotron Q and U emission maps are modelled as power-law and given as an extrapolation in frequency of the polarized 23 GHz {\it WMAP} map:
\begin{equation}
Q_{\nu}(p) = Q_{23}(p)\left(\frac{\nu}{23}\right)^{\beta_s(p)}
\end{equation}
\begin{equation}
U_{\nu}(p) = U_{23}(p)\left(\frac{\nu}{23}\right)^{\beta_s(p)}
\end{equation}
We set the synchrotron spectral index to $\beta_s = -3$ uniformly over the whole sky, consistent with observations by {\it WMAP} \citep{Page07,Gold09}.
The simulated thermal dust Q and U emission maps are also modelled as power-law emission and generated by extrapolating the predicted 94 GHz map in \citet{Fink99}: $S_{\nu}(p) = S_{94}(p)\left(\frac{\nu}{94}\right)^{\beta_d}$.  To generate the dust polarization angles we use a software package called the Planck Sky Model (PSM, version 1.6.6) developed by the {\it Planck} Working Group 2. They closely match the sychrotron angles. The dust polarization fraction is set at 12\%, which is scaled by a geometric depolarization factor due to the expected magnetic field configuration, resulting in an observed polarization fraction of $\sim 4\%$.  We set the dust spectral index to $\beta_d = 1.5$ uniformly over the whole sky. This is consistent with polarization observations by {\it WMAP} at frequencies below 100~GHz, although at higher frequencies thermal emission is observed to deviate from power-law \citep[e.g.,][]{PlanckDust:2011}.

For the parametric model, we assume that the spectral index of the Galactic components do not vary over the frequency range considered, so the coefficients are given by
\begin{align}
\bm\alpha_2(\nu,\beta_2) & = \rm{diag}[(\nu/\nu_{30})^{\bm\beta_2}]  \\
\label{eq:alpha3}
\bm\alpha_3(\nu,\beta_3) & = \rm{diag}[(\nu/\nu_{353})^{\bm\beta_3}].
\end{align}
Here we have defined the two spectral index vectors $\beta_2$ and $\beta_3$ for synchrotron and dust, respectively.  We set the pivot frequencies to 30 GHz and 353 GHz.
We impose Gaussian priors on the spectral index parameters of $\beta_2 = -3.0 \pm 0.3$ for synchrotron and $\beta_3 = 1.5 \pm 0.5$ for dust.  The priors we have chosen have central value and standard deviation at approximately the average and range of values typically observed and predicted theoretically (see, for example, \citet{Fraisse08,Dunkley-cmbpol} for further discussion).

In Test 2 (baseline with non-uniform $\beta_s$), simple power-laws are again used to both simulate the foreground components and also as a model in the separation estimation, but the synchrotron index varies spatially over the sky.  
Dust emission is simulated as in baseline Test 1 but synchrotron emission is modelled as power-law with a spatially varying $\beta_s$. The degree of spatial variation in the polarization spectral index has not yet been well-measured, but a realistic model is taken to be model 4 of \citet{MAMD08}, given by
\begin{equation}
\beta_s = \frac{\log(P_{23}/g f_s S_{408})}{\log(23/0.408)}
\end{equation}
where $P_{23}$ is the {\it WMAP} polarization map at 23 GHz, $g$ is a geometrical reduction factor (reflecting depolarization due to magnetic field structure), $f_s$ is the intrinsic polarization fraction from the cosmic ray energy spectrum, and $S_{408}$ is the 408 MHz map of \citet{Haslam}. The values of $\beta_s$ range from $-3.3$ to $-2.8$.
The parametric model is as described in baseline Test 1, where we fit to power-law synchrotron and dust components. 

\section{Mismatch tests}
\label{sec:mismatch}

The set of tests described below are given a label identifier (A through I) and a short descriptive name to help the reader understand the results.  In each test, we describe the model used to simulate the Galactic foreground component maps (known as the {\it simulation}) and then we describe the model used for the parametric component separation (known as the {\it model}).  The mismatch tests are summarized in Table \ref{tab:mismatch_tests}.  We categorize the mismatch tests into the following three categories: incorrect model (\S\ref{sec:wrong_model}); extra simulated components (\S\ref{sec:extra_comp}); incorrect priors (\S\ref{sec:wrong_priors}).

\begin{table*}
\centering
\begin{tabular}{c l l l}
\hline
{\bf Label} &{\bf Name} & {\bf Simulation} & {\bf Model}  \\
\hline\hline
\multicolumn{4}{|c|}{\bf Baseline Tests} \\
\hline
\multirow{2}{*}{1} & \multirow{2}{*}{Baseline uniform $\beta_s$} & sync power-law $\beta_s$ = -3 & sync power-law $\beta_s = -3 \pm 0.3$\\
& & dust power-law $\beta_d = 1.5$ & dust power-law $\beta_d = 1.5 \pm 0.5$ \\ \hline
\multirow{2}{*}{2} & \multirow{2}{*}{Baseline non-uniform $\beta_s$} & sync power-law $\beta_s = -3.3$ to $-2.8$ & sync power-law $\beta_s = -3 \pm 0.3$\\
& & dust power-law $\beta_d = 1.5$ & dust power-law $\beta_d = 1.5 \pm 0.5$ \\ \hline
\multicolumn{4}{|c|}{\bf Incorrect Model} \\
\hline
\multirow{2}{*}{A} & \multirow{2}{*}{Dust 2-component-a } & sync power-law $\beta_s$ = -3 & sync power-law $\beta_s = -3 \pm 0.3$\\
& & 2-component dust & dust power-law $\beta_d = 1.5 \pm 0.5$ \\ \hline
\multirow{2}{*}{B} & \multirow{2}{*}{Dust 2-component-b} & sync power-law $\beta_s$ = -3 & sync power-law $\beta_s = -3 \pm 0.3$\\
& & 2-component dust & 1-component dust \\ \hline
\multirow{2}{*}{C} & \multirow{2}{*}{Synchrotron curvature} & sync curvature & sync power-law $\beta_s = -3 \pm 0.3$\\
& & dust power-law $\beta_d = 1.5$ & dust power-law $\beta_d = 1.5 \pm 0.5$ \\ \hline
\multicolumn{4}{|c|}{\bf Extra Components} \\
\hline
\multirow{3}{*}{D} & \multirow{3}{*}{1\% Free-free} & sync power-law $\beta_s$ = -3 & sync power-law $\beta_s = -3 \pm 0.3$\\
& & dust power-law $\beta_d = 1.5$ & dust power-law $\beta_d = 1.5 \pm 0.5$ \\ 
& & 1\% polarized free-free & no free-free \\ \hline
\multirow{3}{*}{E} & \multirow{3}{*}{1\% Spinning dust} & sync power-law $\beta_s$ = -3 & sync power-law $\beta_s = -3 \pm 0.3$\\
& & dust power-law $\beta_d = 1.5$ & dust power-law $\beta_d = 1.5 \pm 0.5$ \\ 
& & 1\% polarized spinning dust & no spinning dust \\ \hline
\multicolumn{4}{|c|}{\bf Incorrect Priors} \\
\hline
\multirow{2}{*}{F} & \multirow{2}{*}{Strong $\beta_s$ prior mismatch} & sync power-law $\beta_s = -3.3$ to $-2.8$ & sync power-law $\beta_s = -2.5 \pm 0.5$\\
& & dust power-law $\beta_d = 1.5$ & dust power-law $\beta_d = 1.5 \pm 0.5$ \\ \hline
\multirow{2}{*}{G} & \multirow{2}{*}{Weak $\beta_s$ prior mismatch} & sync power-law $\beta_s = -3.3$ to $-2.8$ & sync power-law $\beta_s = -2.8 \pm 0.5$\\
& & dust power-law $\beta_d = 1.5$ & dust power-law $\beta_d = 1.5 \pm 0.5$ \\ \hline
\multirow{2}{*}{H} & \multirow{2}{*}{Strong $\beta_d$ prior mismatch} & sync power-law $\beta_s = -3$ & sync power-law $\beta_s = -3 \pm 0.3$\\
& & dust power-law $\beta_d = 1.5$ & dust power-law $\beta_d = 2.0 \pm 0.5$ \\ \hline
\multirow{2}{*}{I} & \multirow{2}{*}{Weak $\beta_d$ prior mismatch} & sync power-law $\beta_s = -3$ & sync power-law $\beta_s = -3 \pm 0.3$\\
& & dust power-law $\beta_d = 1.5$ & dust power-law $\beta_d = 1.7 \pm 0.5$ \\ \hline

\end{tabular}
\caption{Summary of mismatch tests.}
\label{tab:mismatch_tests}
\end{table*}

In every case, we define the parametric model for the sky signal using equation \ref{eq:model}.
Given that the CMB radiation is blackbody, the coefficient for $\bm\alpha_1$ is given by $\bm\alpha_1(\nu,\beta_1)  = \bm\alpha_1(\nu) = f(\nu)\mathbf{ I}$, where the function $f(\nu)$ converts the CMB signal $\mathrm{I}$ from thermodynamic to antenna temperature.
Though the spectral indices for Q and U in a given pixel are expected to be similar (following from the assumption that the polarization angle does not change with frequency), unless otherwise stated, we allow the option for the indices to be sampled independently for Q and for U.  Thus, our model is completely described by $6N_p$ amplitude parameters $\mathbf{A} = (\mathbf{A}_{1},\mathbf{A}_2,\mathbf{A}_3)$ and $4N_p$ spectral index parameters $\bm{\beta} = (\bm{\beta}^Q_2,\bm{\beta}^Q_3,\bm{\beta}^U_2,\bm{\beta}^U_3)$.    We impose a flat prior on amplitude-type parameters and Gaussian priors on the spectral index parameters. The model is estimated from $14N_p$ data points (seven frequencies with two Stokes parameters).

We plot the likelihood curves for the estimated parameters, $r$ and $\tau$, for each mismatch case and show the comparison likelihood curve from its corresponding baseline test.  By holding all parameters constant, except for the mismatch being tested, we are able to quantify the level of bias induced by each type of mismatch.  In this section we describe each test and present the numerical results; in Section \ref{sec:discuss} we discuss their implications.

\begin{figure*}  \centering   \includegraphics[width=0.48\textwidth]{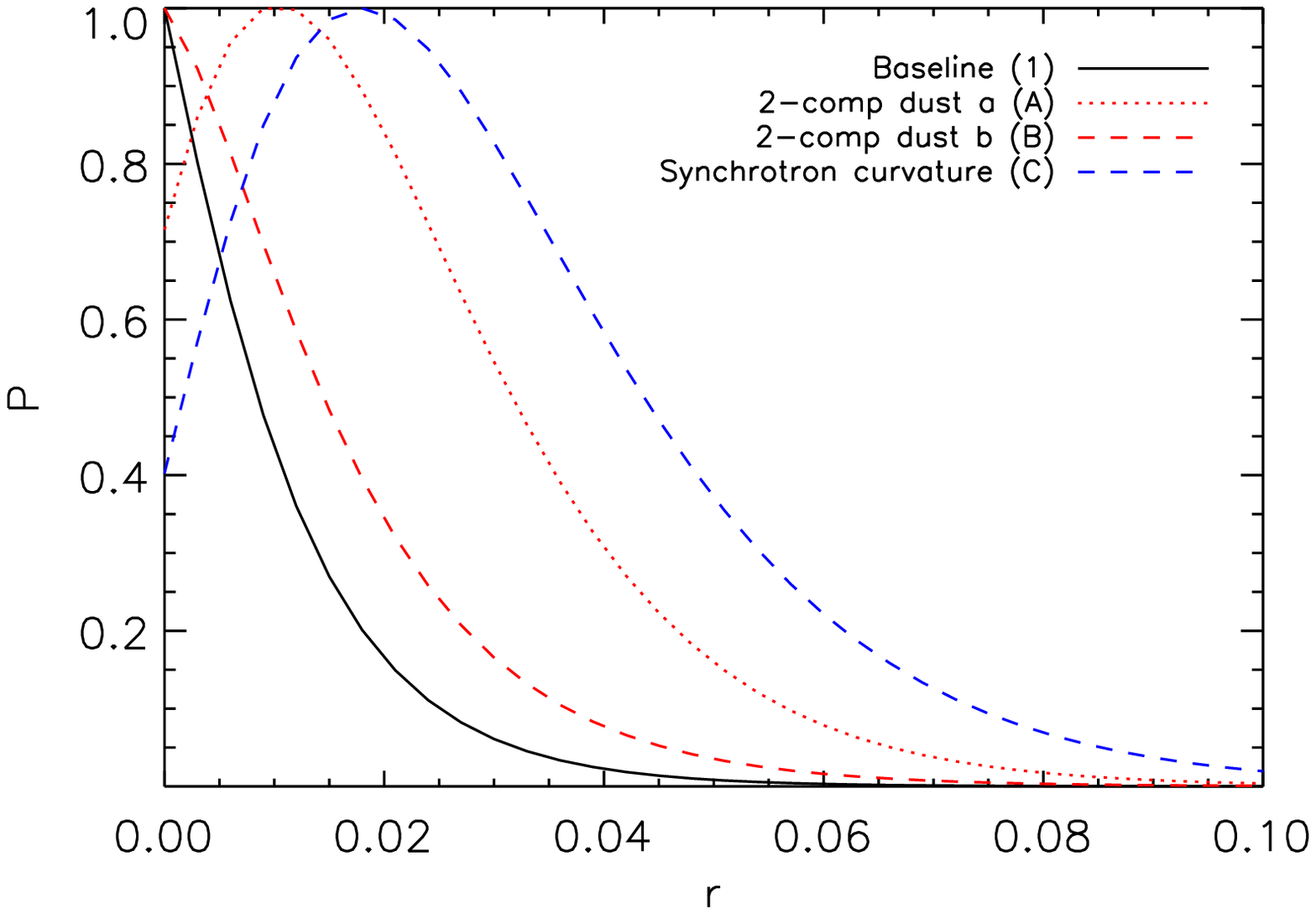} 
     \includegraphics[width=0.48\textwidth]{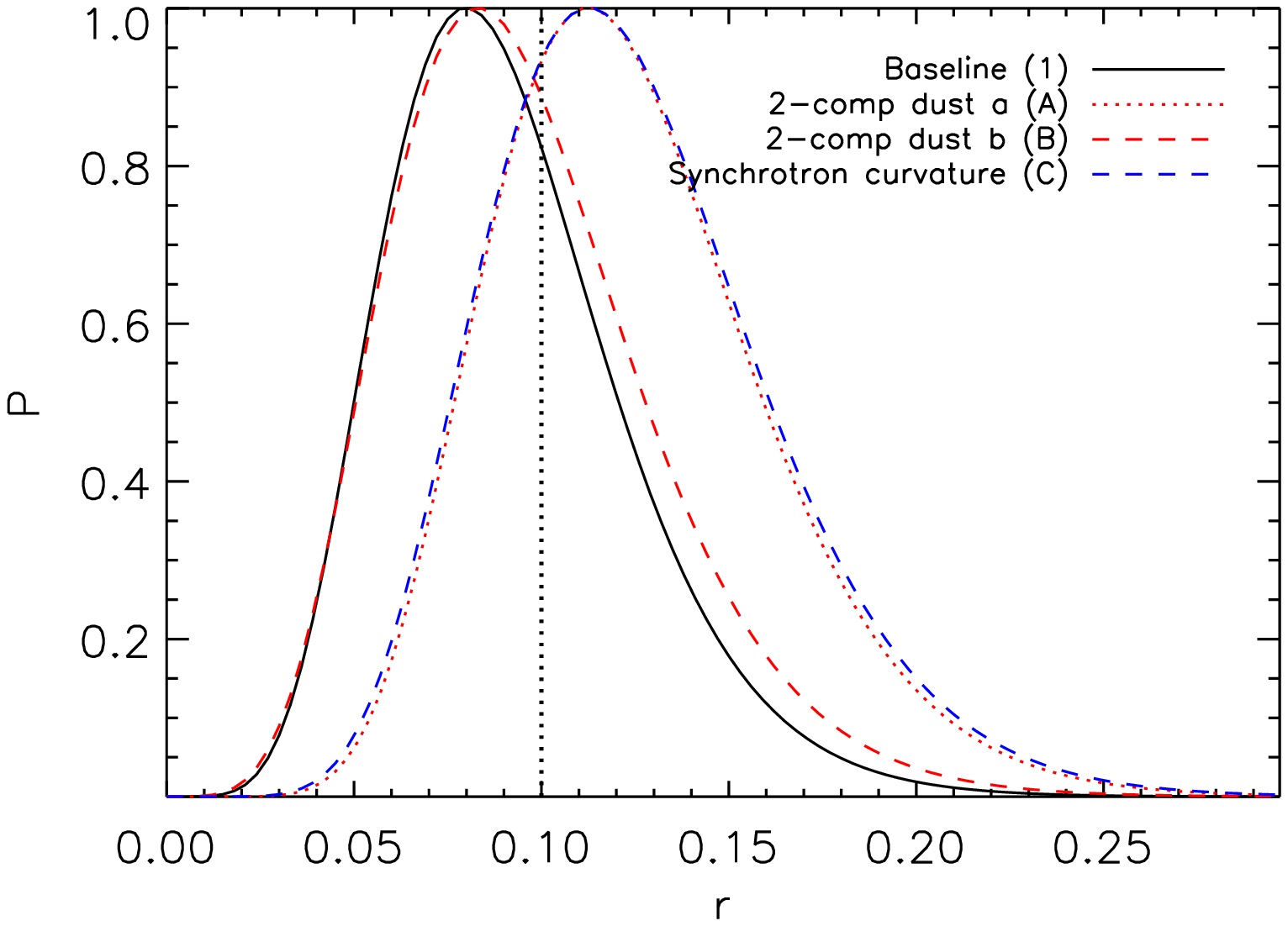} 
   \caption{Recovered distributions for the tensor-to-scalar ratio, $r$, for mismatched simulation and models, comparing the baseline (test 1) with three mismatched cases for $r=0$ (left), and $r=0.1$ (right). Modelling a two-component thermal dust simulation (modified black-body emission with dust at mean temperatures $16$~K and $10$~K) with a power-law dust spectral index (test A) biases $r$ high by about 1$\sigma$, as does neglecting a curvature in the synchrotron spectral index (test C). Modeling a two-component dust simulation with a one-component modified black-body model (Test B) has only a minor effect.}
   \label{fig:wrong_model}
\end{figure*}

\subsection{Incorrect model}
\label{sec:wrong_model}

Here we consider a subset of cases where the frequency dependence of the synchrotron and dust emission are modelled incorrectly. 

\subsubsection{Thermal dust frequency dependence}

Thermal dust emission is well-approximated by a modified black-body, with intensity scaling as $\nu^\beta B_\nu(T)$, where $B_\nu(T)$ is a black-body spectrum with temperature $T$. In the Rayleigh-Jeans limit, this approximates to the power-law assumed in our baseline simulations. Over a broader frequency range, the power-law approximation breaks down, and modelling the curvature becomes important. In the simplest extension to a power law, it is common to fit for one or two parameters to describe the integrated dust emission from any line of sight: either the emissivity index $\beta$, or emissivity plus temperature $T$. More realistically, the integrated dust emission arises from dust grains at various temperatures, so could best be represented by the sum of modified black-bodies. In \citet{Fink99}, a model with just two components at mean temperatures 9.6 K and 16.4~K was found to be a good fit to the IRAS data. 

Here we consider two mismatches between model and simulation. In Test A (two-component-dust-a), the dust emission is simulated with two temperature components, while the parametric model fits to a dust power-law.  We use model 7 of \citet{Fink99}, with  $[Q,U](\nu) \propto A_1 \nu^{\beta_1} B_\nu(T_1) + A_2  \nu^{\beta_2} B_\nu(T_2)$. In this model the first component is sub-dominant, with $A_2/A_1=24.6$.  The dust emissivity indices are $\beta_1=1.5$, $\beta_2=2.6$ over the whole sky. Synchrotron emission is simulated as power-law with a spatially uniform $\beta_s$.  The parametric model fits to power-law dust and synchrotron, neglecting the curvature of the dust spectrum.
In Test B (two-component-dust-b), dust emission is again simulated with two temperature components (as in Test A), while the parametric model fits to a one-component dust model, $[Q,U](\nu) \propto \nu^\beta B_\nu(T)$. We fix the temperature $T$ over the sky to the values of $T_2$ from the simulation, and estimate a single index $\beta_d$ in every pixel.

\begin{table*}
\centering
\begin{tabular}{l c c c c c}
\hline
Test & Recovered $r$ & Recovered $r$ & Bias ($\sigma$) & Recovered $\tau$ & Bias ($\sigma$) \\
& $r=0$ & $r=0.1$ & & $\tau=0.1$ & \\
\hline
\hline
\multicolumn{6}{|c|}{\bf Baseline Tests} \\
\hline
(1) Baseline (uniform $\beta_s$) & $<0.03^{\dag}$ &$0.092\pm0.033$ & --- & $0.094\pm0.005$& -- \\
(2) Baseline (non-uniform $\beta_s$) & $<0.03^{\dag}$ &$0.092\pm0.033$ & --- & $0.094\pm0.005$& -- \\
\hline
\multicolumn{6}{|c|}{\bf Incorrect Model} \\
\hline
(A) Dust 2-component-a & $0.02\pm0.016$ &$0.125\pm0.037$ & $+0.9$ & $0.097\pm0.005$ & $+0.6$\\
(B) Dust 2-component-b & $<0.04^{\dag}$ &$0.096\pm0.036$ & $+0.2$ & $0.094\pm0.005$ & $<+0.1$ \\
(C) Synchrotron curvature & $0.03\pm0.020$ &$0.125\pm0.039$ & $+0.9$ & $0.097\pm0.005$& $+0.6$\\
\hline
\multicolumn{6}{|c|}{\bf Extra Components}  \\
\hline
(D) 1\% free free & $<0.03^{\dag}$ &$0.091\pm0.032$ & $<-0.03$ & $0.094\pm0.005$  & $<+0.1$\\
(E) 1\% spinning dust & $<0.04^{\dag}$ &$0.094\pm0.033$ & $<+0.03$& $0.094\pm0.005$ & $<-0.1$\\
\hline
\multicolumn{6}{|c|}{\bf Incorrect Priors} \\
\hline
(F) Strong $\beta_s$ prior mismatch & $0.168\pm0.047$ &$0.197\pm0.047$ & $+2.1$ & $0.104\pm0.006$ & +1.7\\
(G) Weak $\beta_s$ prior mismatch & $0.029\pm0.021$ &$0.117\pm0.039$ & $+0.6$ & $0.096\pm0.005$ & +0.4\\
(H) Strong $\beta_d$ prior mismatch & $0.133\pm0.044$ &$0.224\pm0.040$ & $+3.3$ &$0.107\pm0.005$ & +2.6\\
(I) Weak $\beta_d$ prior mismatch & $<0.04^{\dag}$ &$0.111\pm0.034$ & $+0.6$ & $0.096\pm0.005$ & +0.4\\
\hline
\end{tabular}
\caption{Marginalized estimates and corresponding biases for $r$ for simulations with $r=0$ and $r=0.1$, and for $\tau$ for simulations with $\tau=0.1$. $^{\dag}$These values are the upper $95\%$ confidence levels for $r=0$.}
\label{tab:recovered_r}
\end{table*}

Using these test cases, we perform component separation and use the resulting CMB maps to compute the likelihoods for parameters $\tau$ and $r$ for the $r=0$ and $r=0.1$ simulations.  
The distributions are shown in Fig.~\ref{fig:wrong_model}, and recovered mean values for $r$, and $\tau$, for these and all other tests are summarized in Table \ref{tab:recovered_r}. For $r=0$ we quote 95\% upper limits; for $r=0.1$ and $\tau$ we give 68\% confidence levels. For $r=0.1$ we find a non-negligible bias on $r$ of $1\sigma$ high for Test A, fitting a two-component dust model with a power-law, and a similar bias high for the optical depth, $\tau$. Using a one-temperature component model to fit the two-component simulation (Test B), recovers $r$ with only $\sim 0.2\sigma$ bias. We see a similar effect for the $r=0$ case, where for Test A the recovered $r$ value is greater than zero at $1\sigma$, but Test B is consistent with the baseline case.

\subsubsection{Synchrotron frequency dependence}

Synchrotron emission is expected to be roughly power-law in frequency \citep[see e.g.,][]{RL79}, the result of relativistic cosmic-ray electrons 
accelerated in the Galactic magnetic field \citep{strong/moskalenko/ptuskin:2007}. However, a steepening of the index with frequency is also expected, due to increased energy loss of the electrons \citep[e.g.,][]{banday/wolfendale:1991,strong/moskalenko/ptuskin:2007}. The {\it WMAP} data are consistent with power-law emission, but a modest steepening would fit the data, and can be parameterized by a curvature of the spectral index. In a pessismistic scenario, the degree of steepening could vary significantly over the sky, or the frequency dependence could be ill-fit by a single curvature parameter.

In Test C (synchrotron curvature), the simulated Galactic foreground includes a steepening of the synchrotron index with frequency while the parametric model retains power-law synchrotron emission.  The synchrotron emission has spectral curvature such that the index decreases by $0.3$ above 23 GHz.
 Figure \ref{fig:wrong_model} and Table \ref{tab:recovered_r} show the results from this third test case. The effect on the recovered CMB is non-negligible. We find that a synchrotron curvature simulation generates a bias of about $1\sigma$ high in $r$, or $\delta r \sim 0.03$, roughly the same level as the two-component dust simulation with power-law model. This mismatch also results in a 1.5$\sigma$ preference for $r>0$ for the $r=0$ model.

\begin{figure*}  \centering  \includegraphics[width=0.48\textwidth]{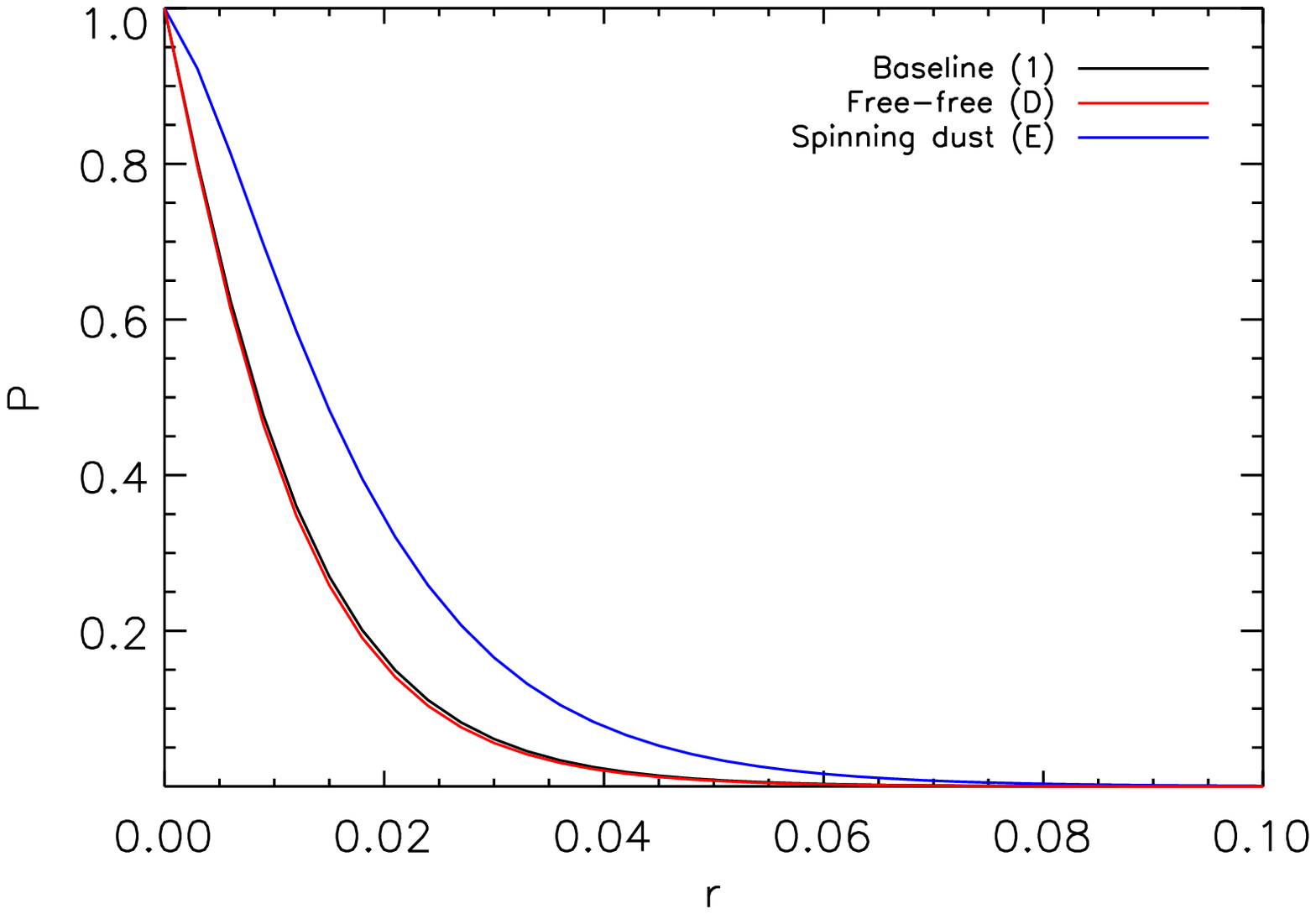}    \includegraphics[width=0.48\textwidth]{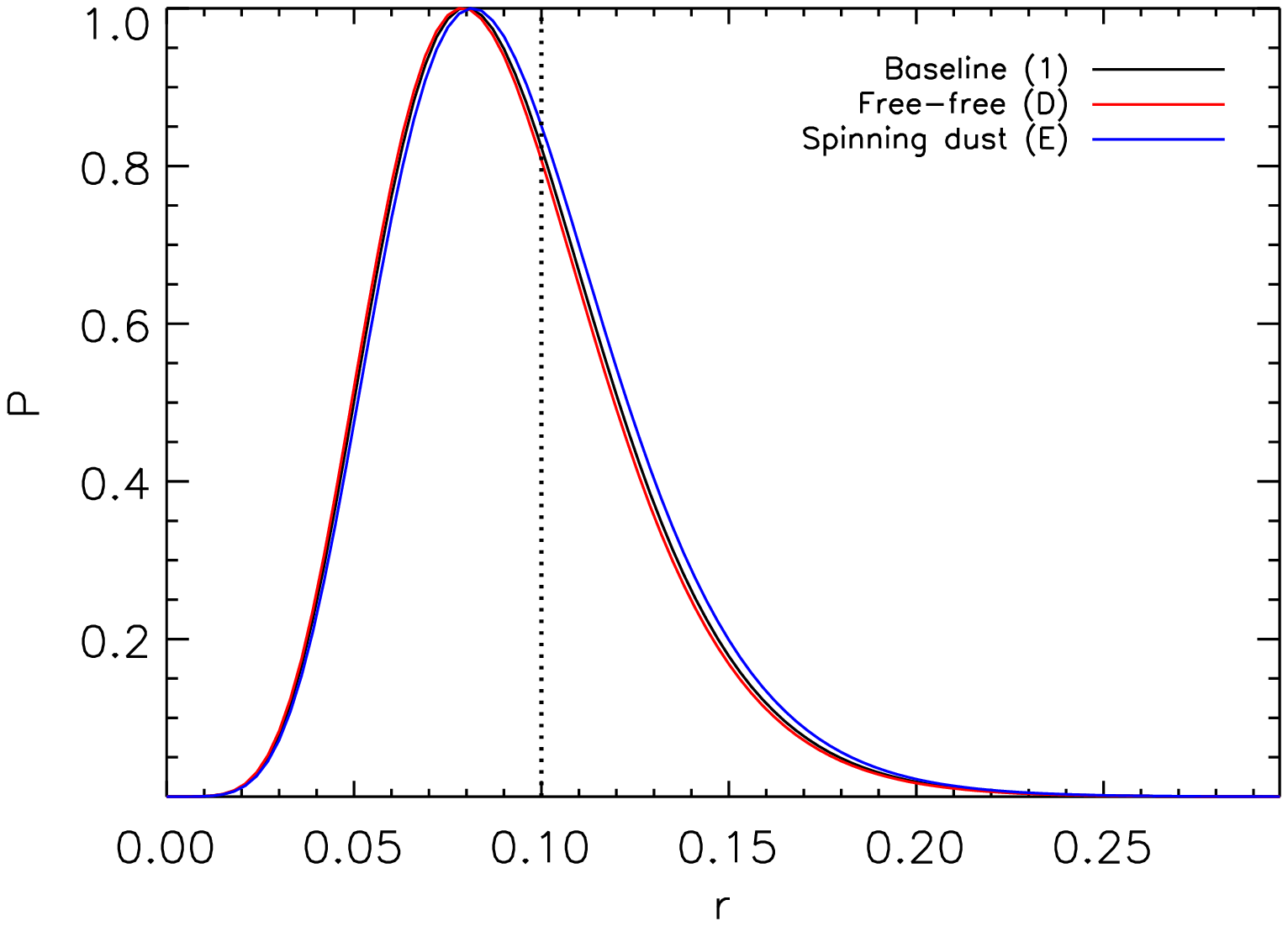}
   \caption{Recovered distributions for the tensor-to-scalar ratio, $r$, for simulations containing polarized components that are neglected in the models. The baseline results (test 1) are compared to those with a 1\% polarized
free-free component (test D), and a 1\% polarized spinning dust component (test E), for $r=0$ (left), and $r=0.1$ (right). At this polarization level, these components are sufficiently sub-dominant that they do not bias the recovered parameters.}
\label{fig:extra_components}
\end{figure*}

\subsection{Additional polarized components}
\label{sec:extra_comp}

Our model and simulations contain only synchrotron and thermal dust emission components. Other emission components are not expected to be significantly polarized (see e.g., \citet{Fraisse08}, and Section \ref{sec:discuss} for further discussion). However, both free-free and spinning dust emission are detected in intensity, and they may be minimally polarized at the few-percent level. \citet{Macellari11} find an upper limit on spinning dust of $5\%$ and an upper limit on free-free polarization of $<3\%$. \citet{Dickinson11,Lopez-Caraballo11} reduce the upper limits on spinning dust polarization to $\sim 1-2\%$.

Test D (free-free) simulates a Galactic foreground that includes a 1\% polarized free-free emission in addition to the synchrotron and dust emission. Free-free Q and U emission are given by $Q_{\rm ff}(\nu) = 0.01 I_{\rm ff}(\nu) \cos(2\gamma)$ and $U_{\rm ff}(\nu) = 0.01 I_{\rm ff}(\nu) \sin(2\gamma)$, where $I_{\rm ff}(\nu)$ is a free-free intensity map at frequency $\nu$ and $\gamma$ are the thermal dust angles. This assumes that the free-free polarization angles match the thermal dust angles, which is unrealistic but should not significantly affect conclusions. The free-free intensity is generated from the PSM, which is consistent with {\it WMAP} data. The parametric model fits for power-law synchrotron and dust but omits the free-free component.

Test E (spinning dust) includes a 1\% polarized spinning dust emission in addition to synchrotron and thermal dust. Spinning dust Q and U emission are given by $Q_{\rm sd}(\nu) = 0.01 I_{\rm sd}(\nu) \cos(2\gamma)$ and $U_{\rm sd}(\nu) = 0.01 I_{\rm sd}(\nu) \sin(2\gamma)$, where $I_{\rm sd}(\nu)$ is a spinning dust intensity map at frequency $\nu$ estimated from the PSM, and the angles $\gamma$ are  the same as the thermal dust angles. The parametric model omits the spinning dust component.

 The resulting likelihoods are shown in Fig.~\ref{fig:extra_components}, and parameters given in Table \ref{tab:recovered_r}. We find that these small unmodelled components have a negligible effect on the estimated parameters; the induced biases are within $0.04\sigma$ of the baseline measurement in each case.

\subsection{Incorrect priors}
\label{sec:wrong_priors}

\begin{figure*}
  \centering
  \includegraphics[width=0.45\textwidth]{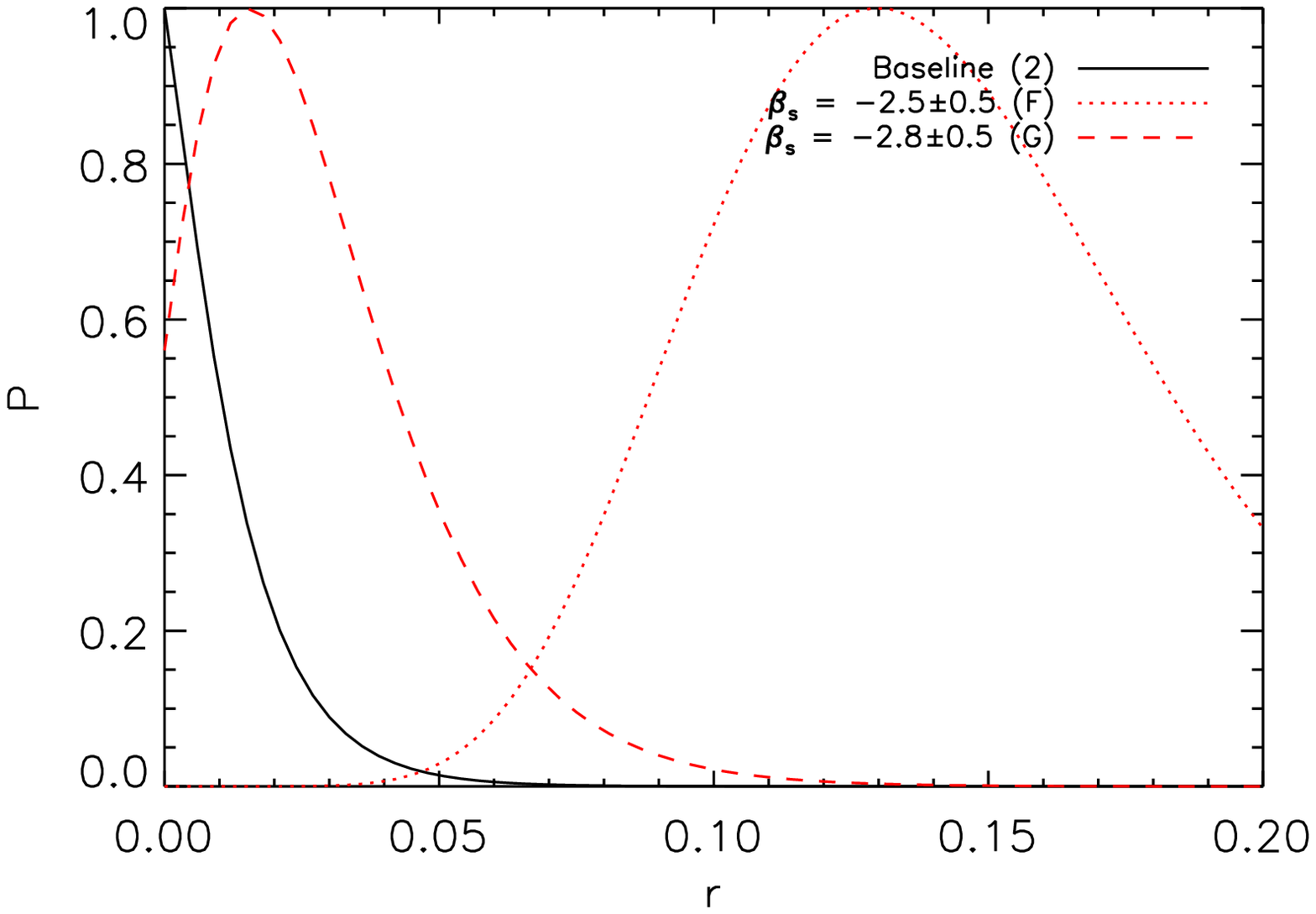}
   \includegraphics[width=0.45\textwidth]{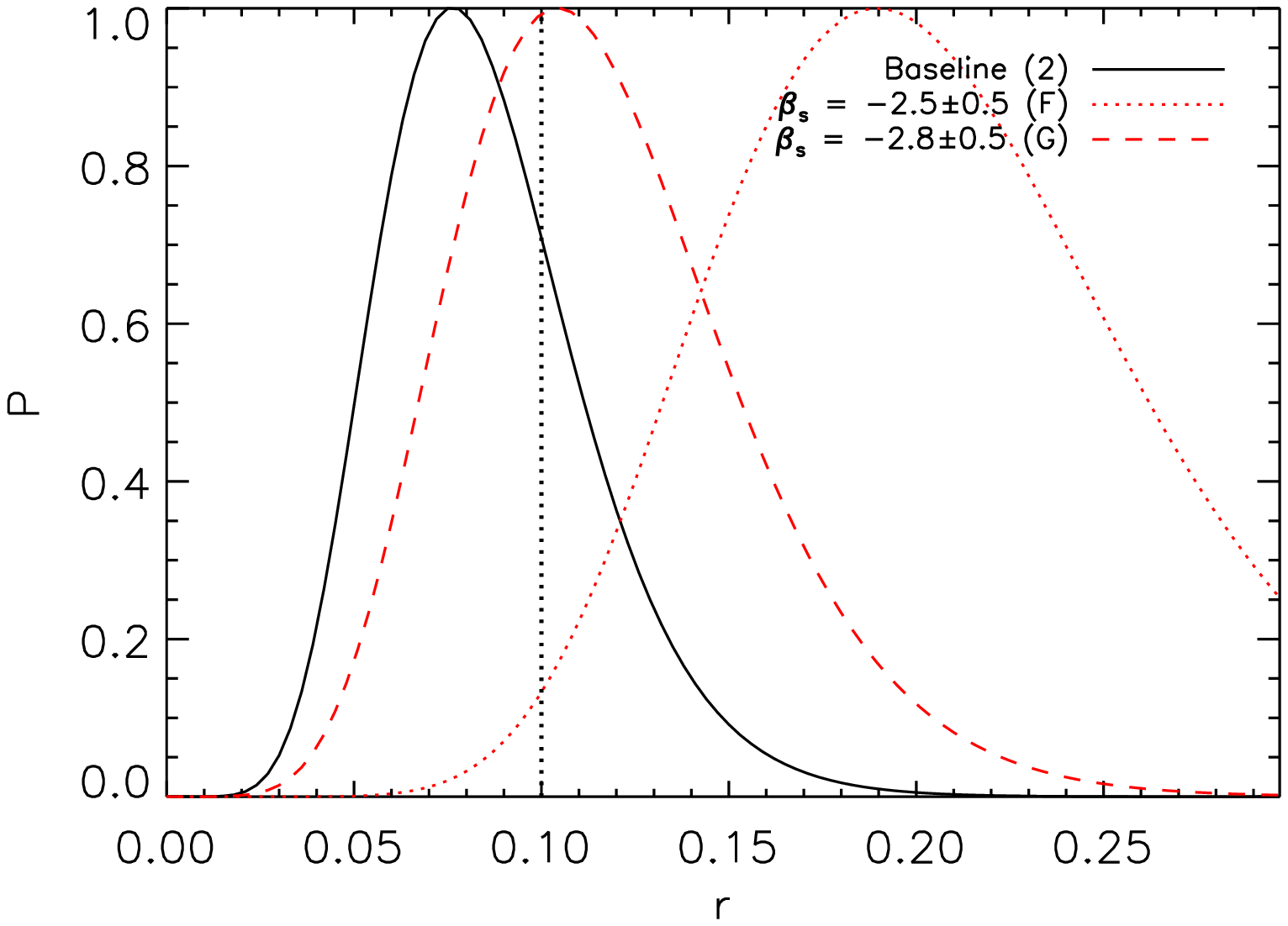}
  \includegraphics[width=0.45\textwidth]{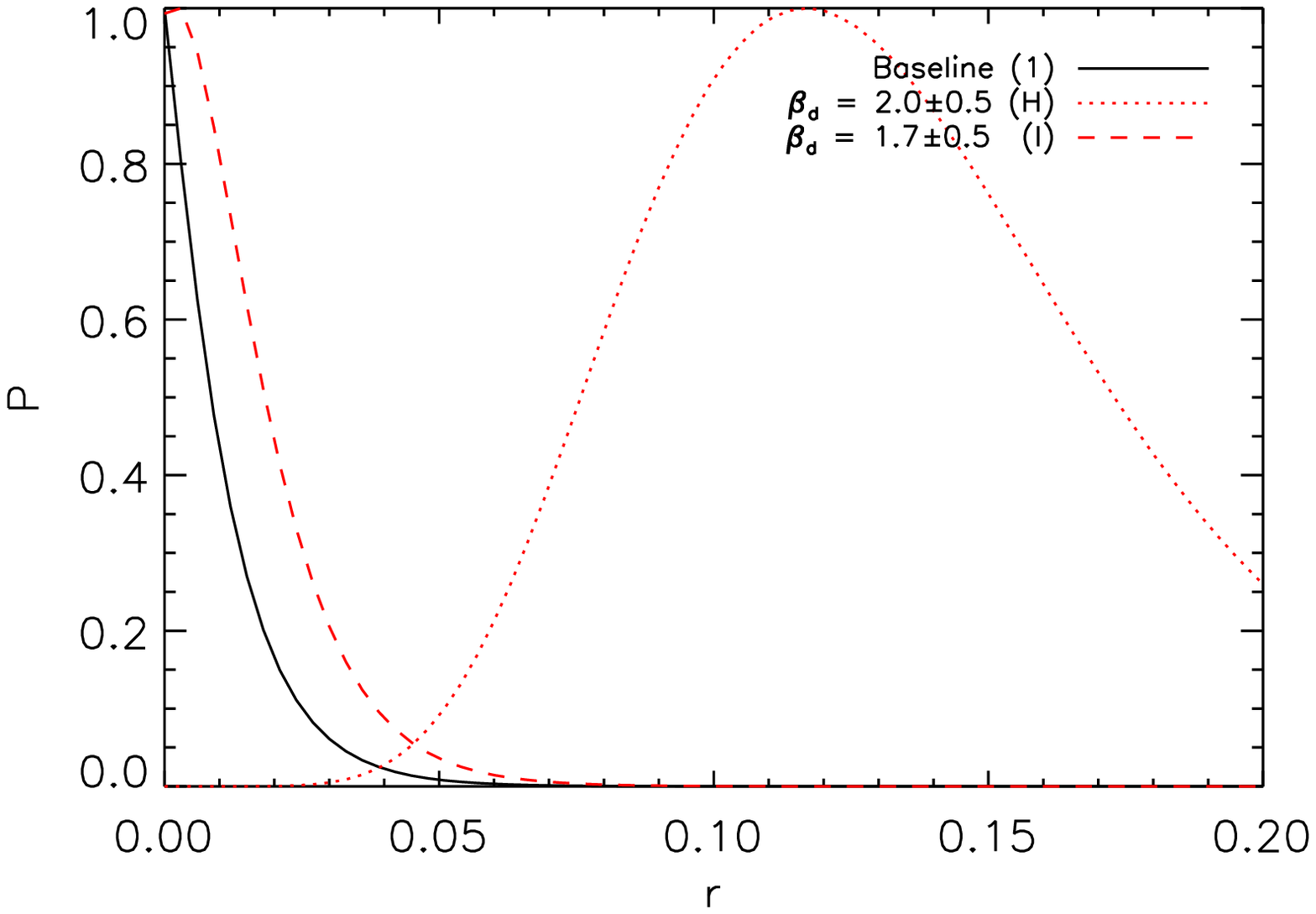}
   \includegraphics[width=0.45\textwidth]{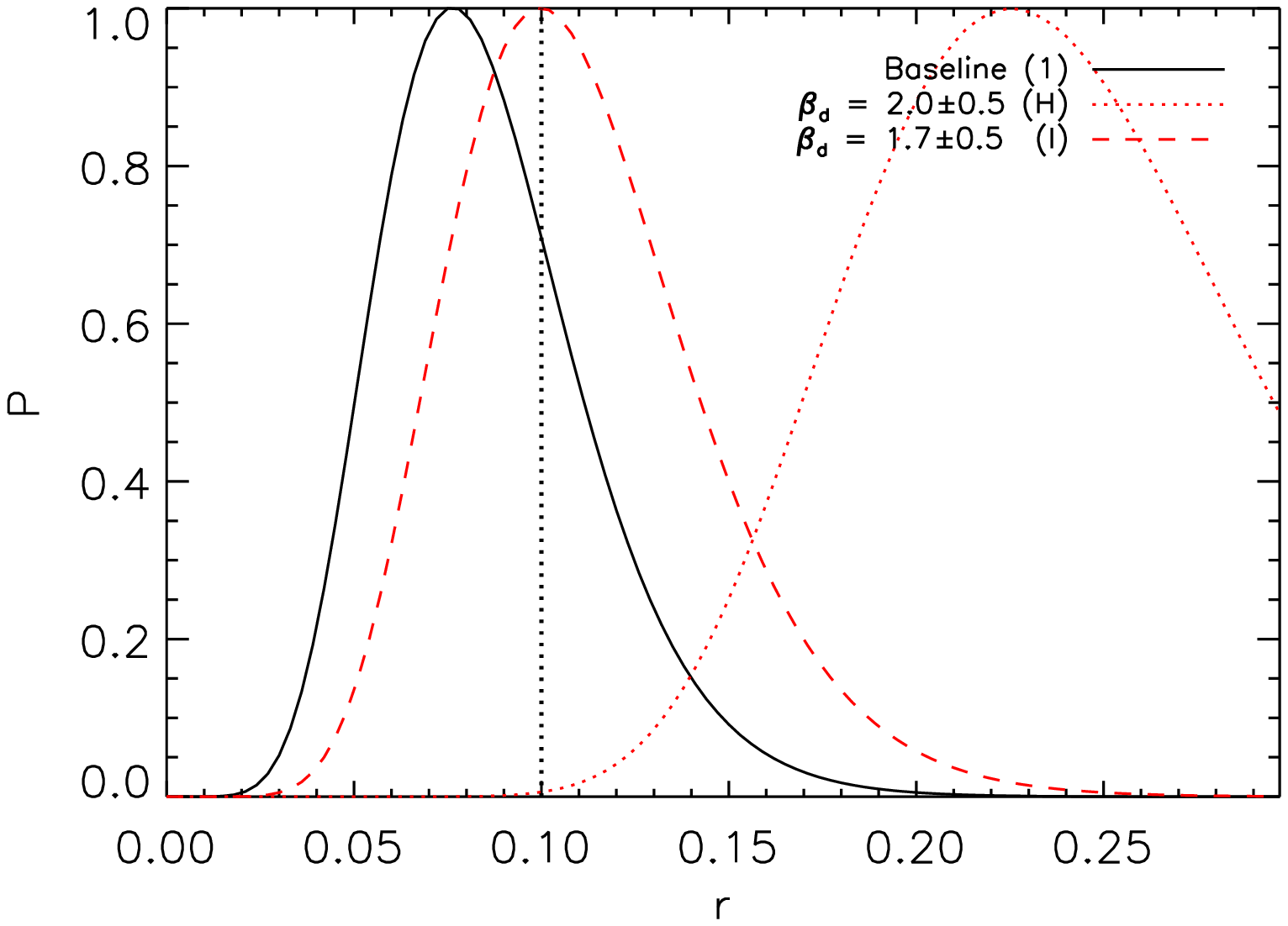}
   \caption{Recovered distributions for $r$, if prior distributions are imposed on spectral indices that do not exactly match the
simulation inputs.  The baseline (test 2) has a dust index $\beta_d=1.5$, and a synchrotron index with mean $\beta_s=-3$ over the sky. Indices for Stokes Q and U are fit in 3-degree pixels over the sky, with Gaussian priors $\beta_s=-3\pm0.3$ and $\beta_d=1.5\pm0.5$. Offsetting the synchotron prior by 1$\sigma$ to $-2.5\pm0.5$ (test F), significantly biases the recovered $r$ high (top panels, for $r=0$, left,  and $r=0.1$, right).  A $\sim 0.5\sigma$ offset (test G) results in a smaller but non-negligible bias. Similar biases are found for offsets in the dust prior (bottom), for $\beta_d=2.0\pm0.5$ (test H) and $\beta_d=1.7\pm0.5$ (test I). These biases arise from over-parameterizing the model in low signal-to-noise regions.}
\label{fig:betad_prior}
\end{figure*}

In our baseline model estimation we imposed Gaussian priors of $\beta_s=-3\pm0.3$ for the synchrotron spectral index, and $\beta_d=1.5\pm0.5$ for the thermal dust emissivity index. This allowed an estimate of the CMB in areas of the sky with a low signal-to-noise ratio. Even with seven frequencies, if the signal-to-noise ratio is low, the synchrotron and dust component can become degenerate with the CMB unless priors are imposed. 

The priors are astrophysically motivated; synchrotron emission is expected to have an index in the typical range $-3.5 \ltsim \beta_s \ltsim -2.5$, depending on the injection spectrum and nature of diffusion and cooling \citep{RL79,Fraisse08}. Thermal dust emission is expected to have emissivity index in the range $1\ltsim\beta\ltsim2.5$ \citep[see e.g.,][]{Fraisse08}. The 2$\sigma$ range of the prior therefore captures physically reasonable beheaviour. However, our simulations are perfectly matched to these priors: the simulated synchrotron indices are either exactly $-3.0$ in Test 1, or have a mean over the sky of $-3$ in Test 2, and the dust was simulated to have an index of $1.5$. The real sky will likely not match so well: we expect the emission to lie in the prior range, but will not precisely match the mean.  \citet{Dickinson09} conducted a similar study to quantify the effect of priors using real data.  Though they found that the priors had a small impact on the CMB spectra, they considered unpolarized emission, where foregrounds are relatively smaller. 

We test the effects of these prior choices by fixing the simulation spectral behavior, but choosing alternative Gaussian priors with means that are offset from the simulation inputs.

Test F (`strong' $\beta_s$ prior mismatch) examines a reasonably strong case of mismatch between the model prior and simulation for synchrotron.  Using Test 2 as the baseline, it simulates synchrotron emission with values of $\beta_s$ that range between $-3.3$ and $-2.8$, but the parametric model assumes power-law synchrotron with a prior on $\beta_s$ of $-2.5\pm0.5$. Test G (`weak' $\beta_s$ prior mismatch) assumes a prior of $-2.8\pm0.5$. Test H (strong $\beta_d$ prior mismatch) has a mismatch between the model prior and simulation for dust.  Using the baseline simulations, the dust emission has $\beta_d = 1.5$ while the parametric model assumes a prior on $\beta_d$ of $2.0\pm 0.5$. Test I (`weak' $\beta_d$ prior mismatch) assumes a prior of $1.7\pm 0.5$. 

 The likelihoods for these cases are plotted in Fig.~\ref{fig:betad_prior}, with parameters reported in Table \ref{tab:recovered_r}. These mismatches result in the most significant biases. For synchrotron, the strong mismatch case results in a 3.5$\sigma$ spurious detection of $r$ ($0.17\pm0.05$), for a model with no tensor component. The recovered value for $r$ is also biased about $2\sigma$ high for the $r=0.1$ case, and the optical depth $\tau$ is high by almost $2\sigma$. 
The weak mismatch case, with prior $-2.8\pm0.5$, is biased by $\sim0.6\sigma$ in $r$, with a spurious signal at the 1$\sigma$ level.  Similar results are seen for the dust emission. For the strong mismatch a signal is significantly detected at $3\sigma$ when $r=0$, and biased more than $3\sigma$ for $r=0.1$ (returning $0.22\pm0.04$). The weak mismatch case suffers from a bias of $0.6\sigma$ in $r$, and $0.4\sigma$ in $\tau$.

\section{Discussion}
\label{sec:discuss}

We have found that modelling polarized Galactic foregrounds incorrectly can lead to significant biases in the recovered CMB signal. In this section we discuss the reasons these biases are observed, and how they might be mitigated.

\subsection{Effect of priors}
\label{subsec:priors}

\begin{figure*}
   \begin{tabular}{ccc}
      \includegraphics[width=0.2\textwidth,angle=90]{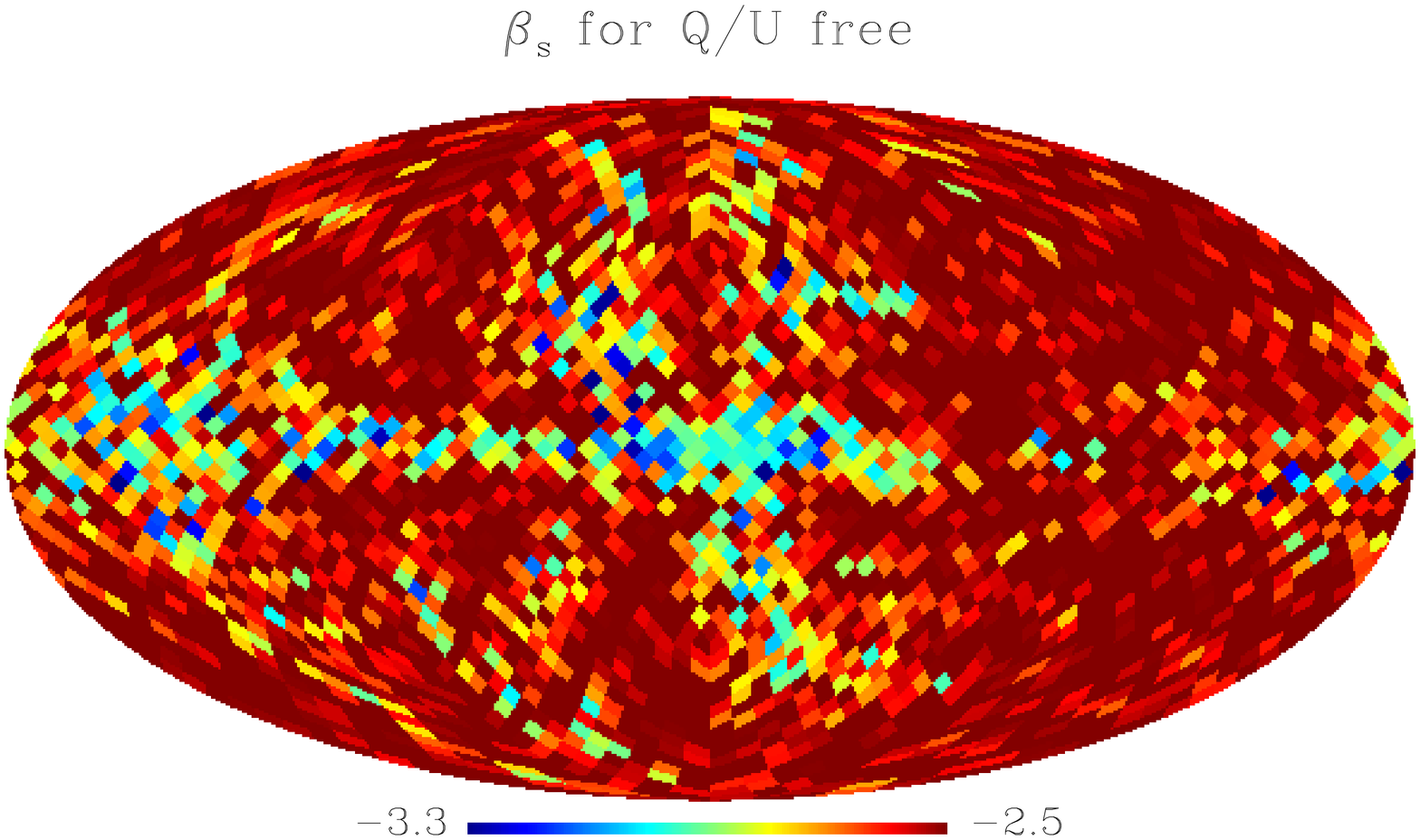} &
  \includegraphics[width=0.2\textwidth,angle=90]{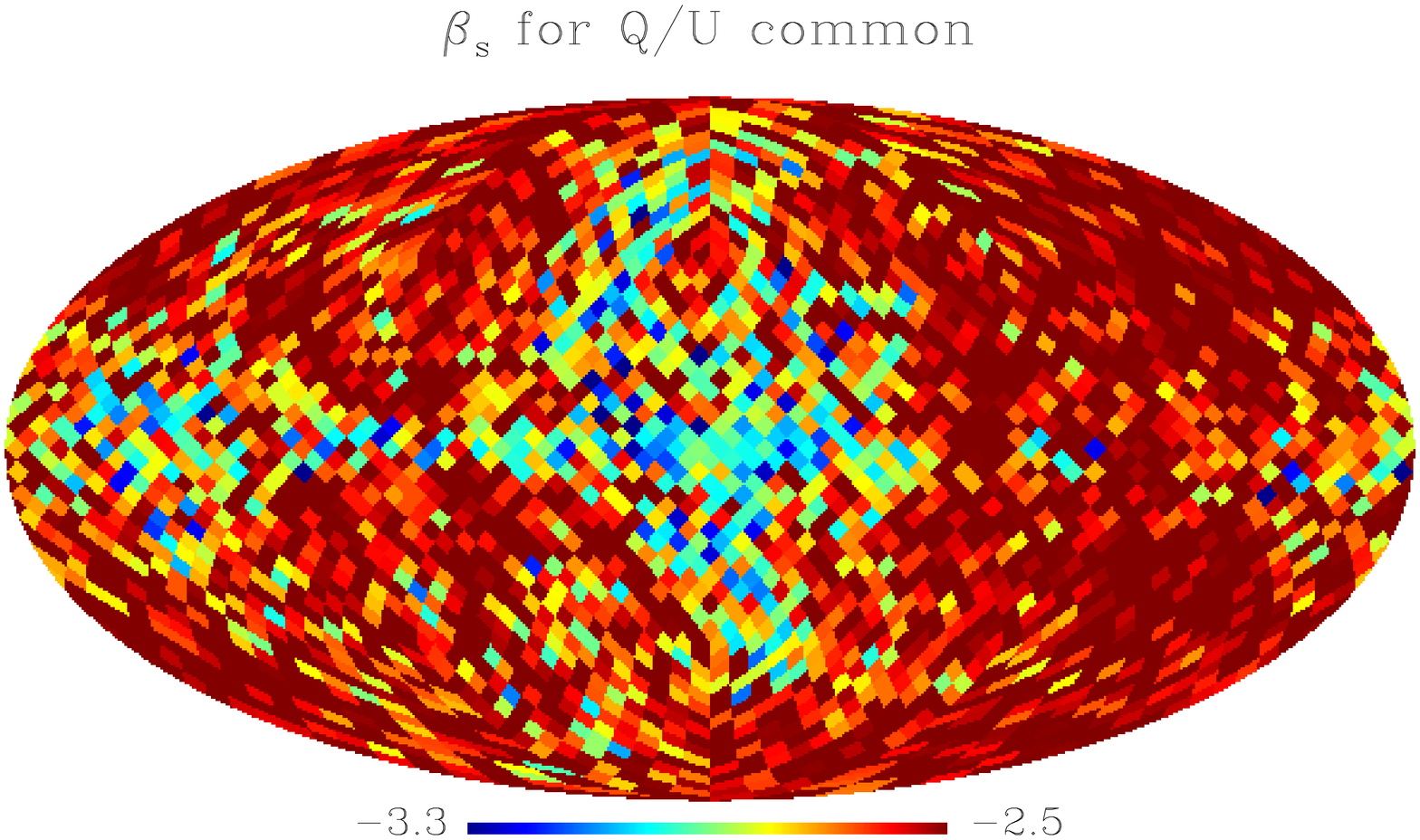}&
      \includegraphics[width=0.2\textwidth,angle=90]{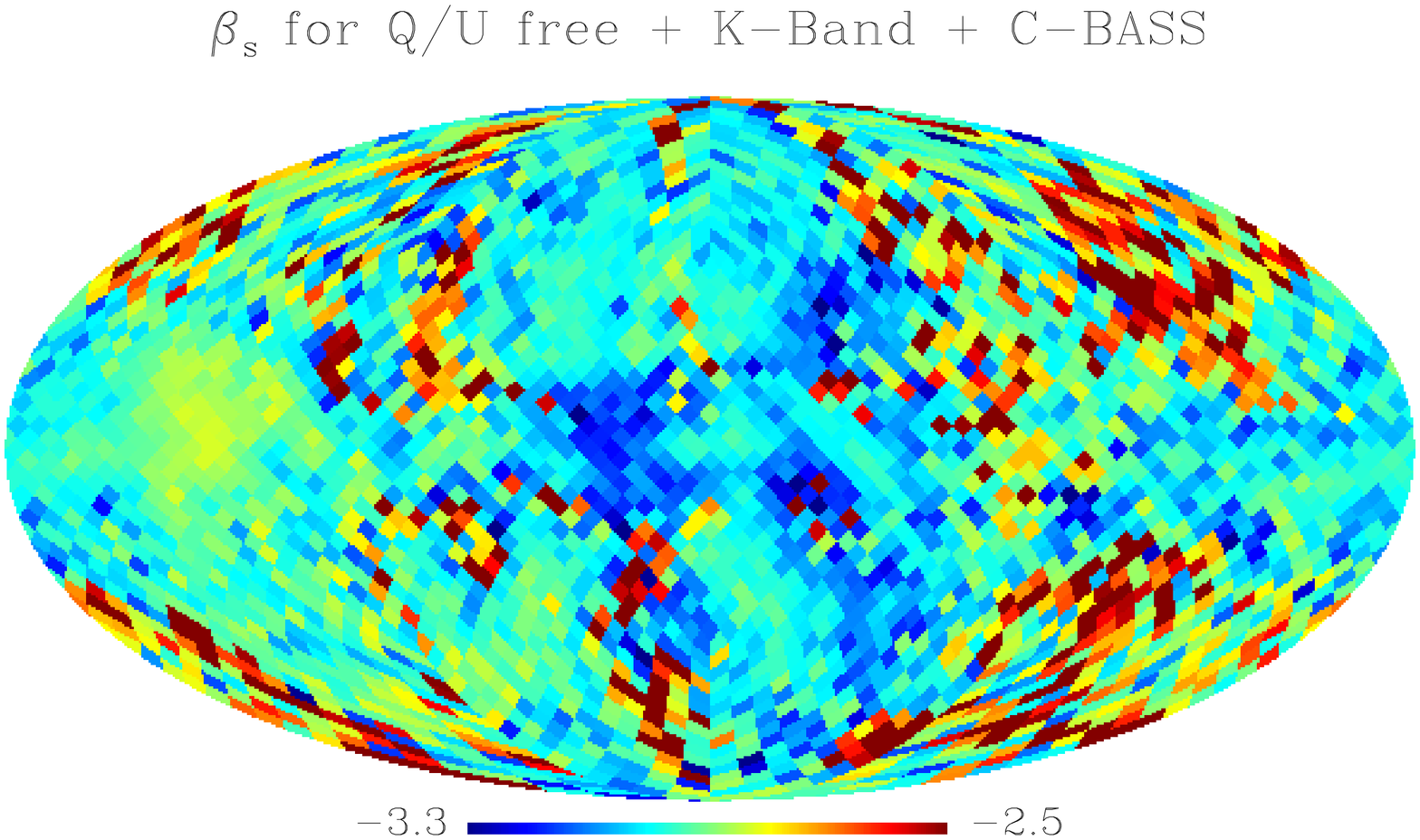}  \\

   \includegraphics[width=0.2\textwidth,angle=90]{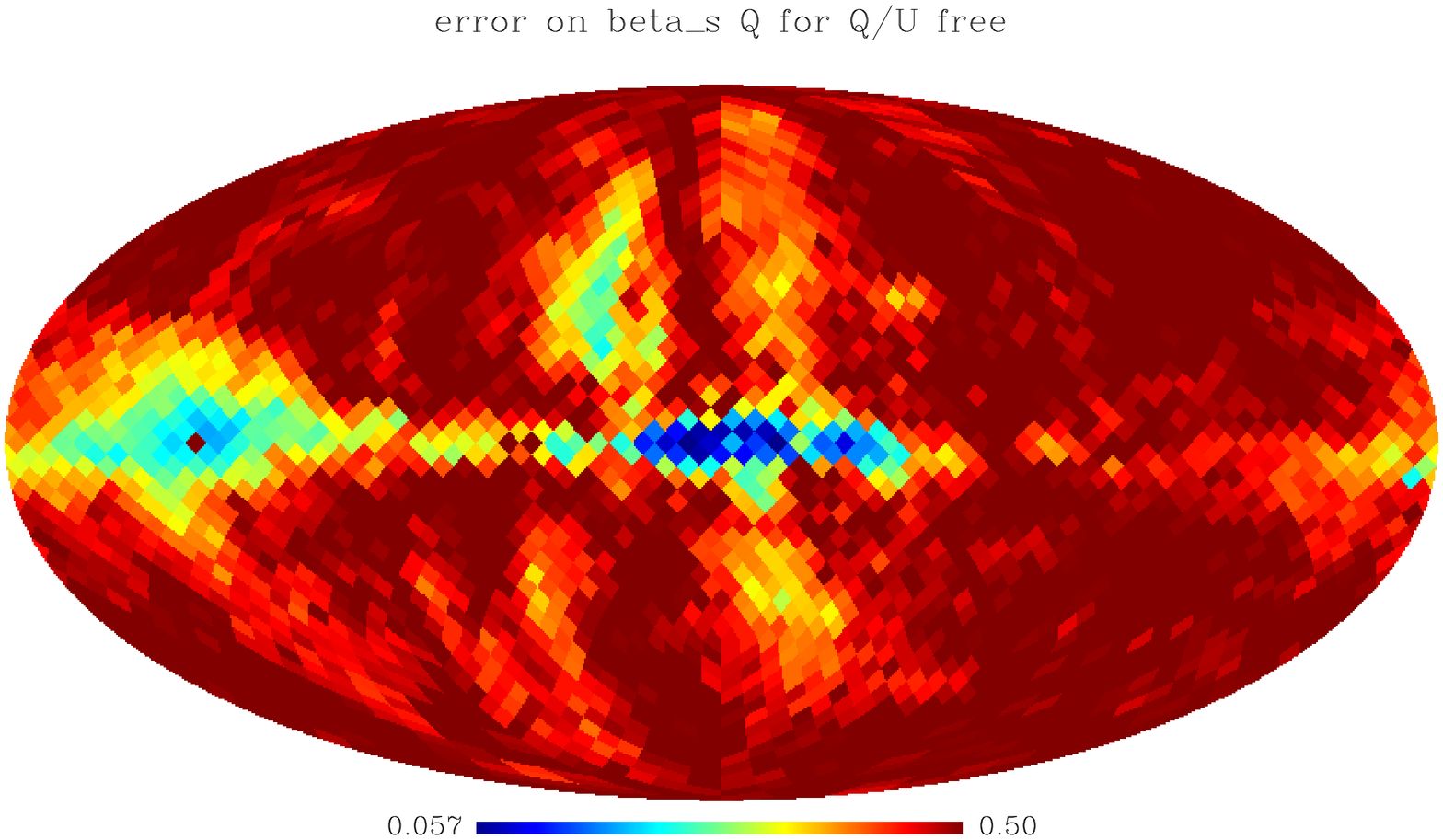} &
  \includegraphics[width=0.2\textwidth,angle=90]{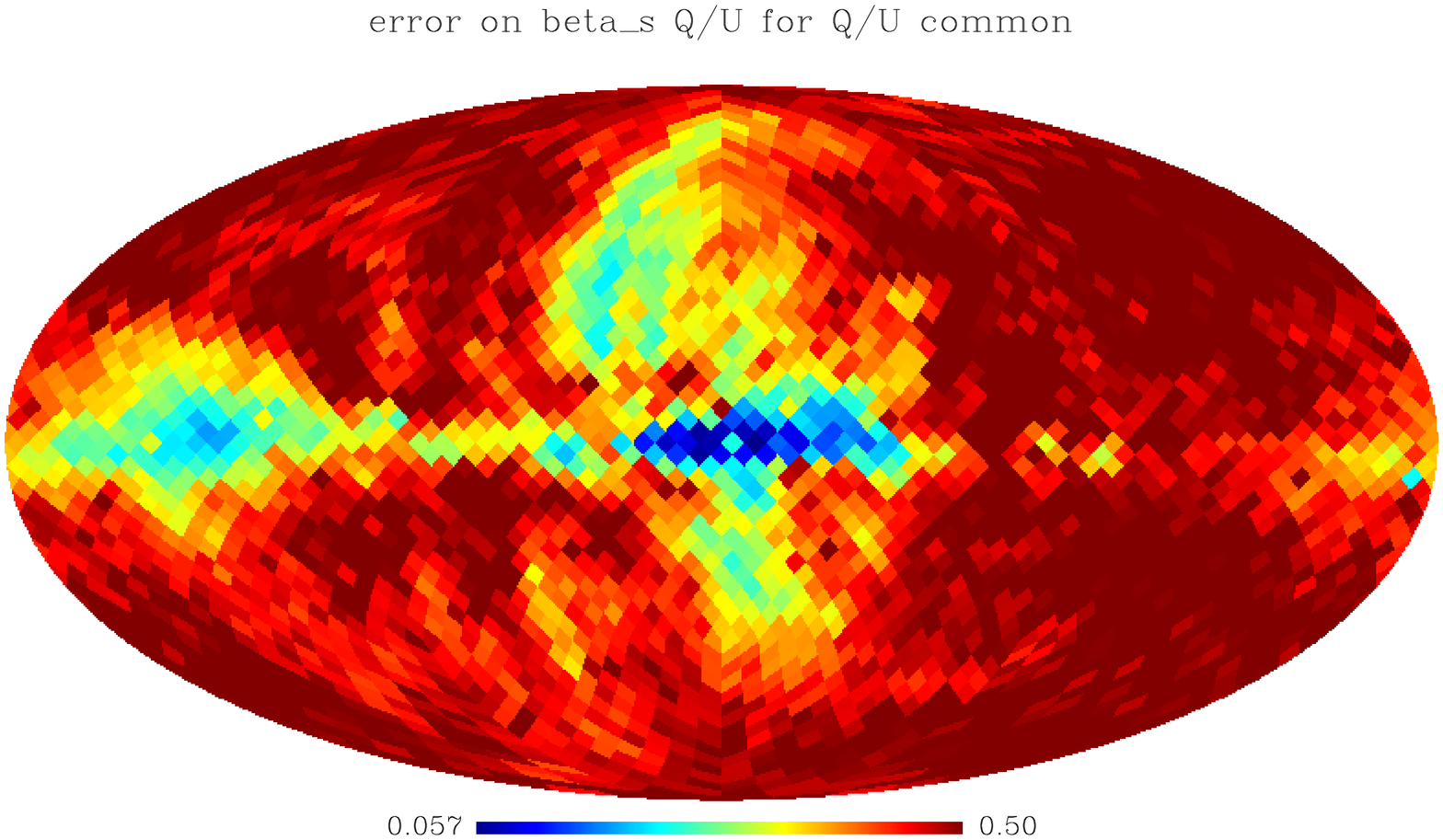}&
      \includegraphics[width=0.2\textwidth,angle=90]{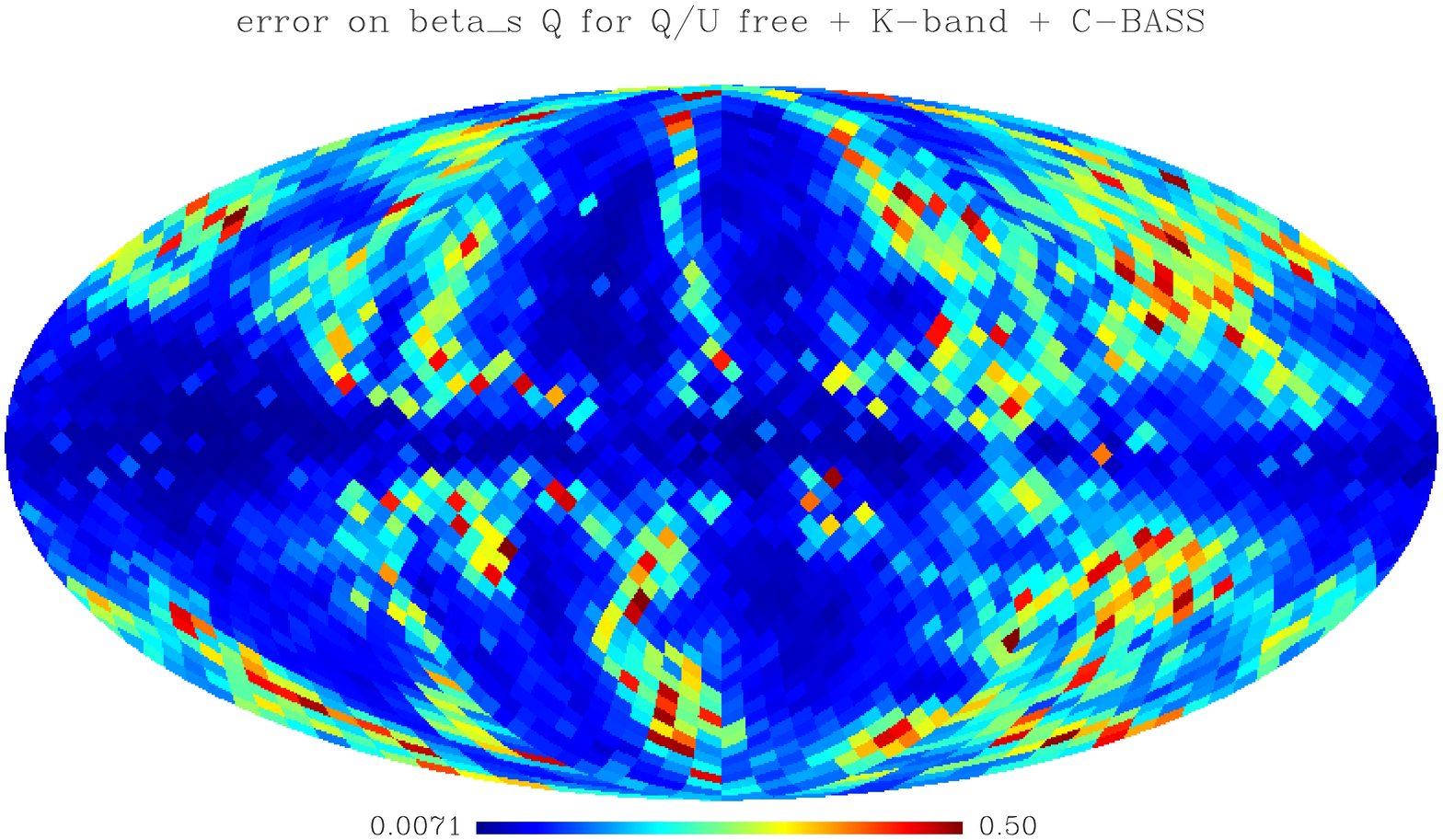}  
      \end{tabular}
      \caption{Estimated synchrotron spectral index for the Q-Stokes parameter (showing mean, top, and uncertainty, bottom), for a simulation with mean $\beta_s=-3$ and prior $-2.5\pm0.5$ (Test F). Allowing free Q and U spectral indices, and using just 30-353~GHz data (left), the prior of $-2.5$ is returned in low signal-to-noise regions. If Q and U signals are assigned a common index (centre), the signal-to-noise is increased. If low-frequency simulated data from {\it WMAP} (23~GHz) and C-BASS (5~GHZ) is added (right), the spectral index map is recovered with high signal-to-noise.}
\label{fig:betas_maps}
\end{figure*}

When marginalizing over foreground uncertainty using a parameterized method, components are distinguished by their frequency dependence. This provides a way of separating the black-body CMB signal from the foreground components.  In the low signal-to-noise regime a prior on this spectral behavior breaks the degeneracy between CMB and foregrounds.

However, we find that choosing an incorrect, yet physically reasonable, prior for the frequency dependence can have a significant impact on the estimated cosmological signal. With a simulated synchrotron spectral index between $-3.3$ and $-2.8$, and a Gaussian prior of $-2.5\pm0.5$ on the index in each pixel, the tensor-to-scalar ratio is overestimated by $\sim 3\sigma$ for an $r=0.1$ model, or a spurious detection made when $r=0$. The effect is less extreme when the mean of the Gaussian prior is closer to the input, $-2.8$, but a bias of 1$\sigma$ is still observed. In the limit of a low signal-to-noise ratio, this can be understood as equivalent to setting the spectral index to the wrong value over the whole sky. A prior of $\beta_s=-2.5\pm0.5$ results in an index that is everywhere $\sim-2.5$, instead of the mean simulated value $\beta_s\sim -3$. Similarly, a prior on the dust index, or emissivity, of $\beta_d=2.0\pm0.5$ results in an index of $\sim 2.0$ instead of the simulated $1.5$. 

\begin{figure*}
     \includegraphics[width=0.8\textwidth,angle=0]{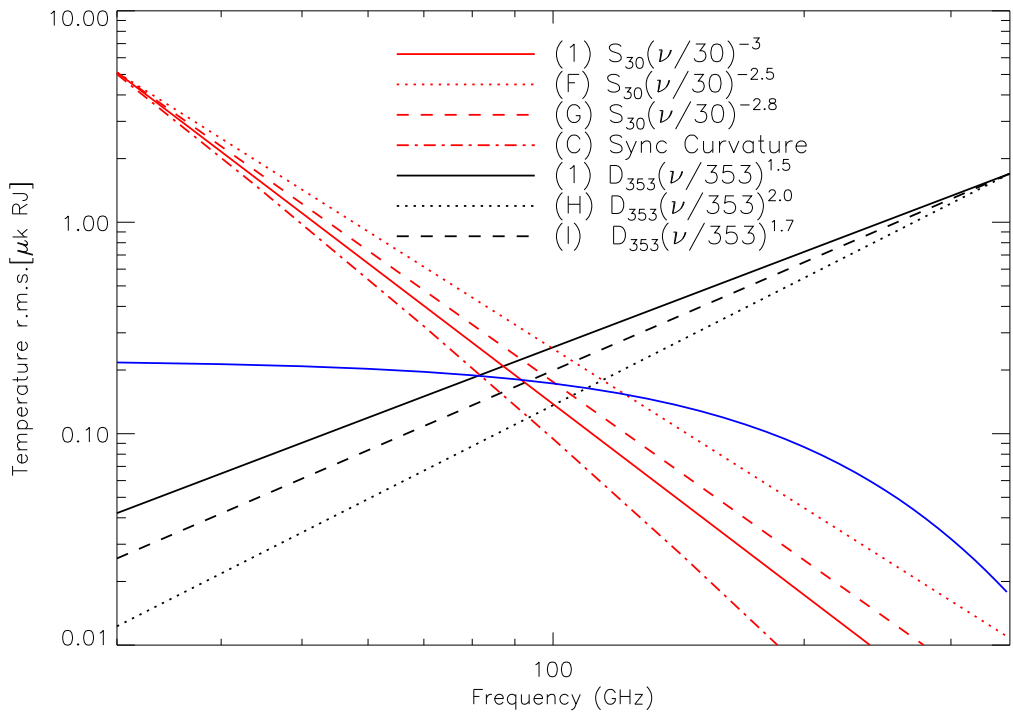}
      \caption{Frequency scaling of the foreground components in the baseline simulation (test 1), rms averaged over the unmasked sky in $3.7^\circ$ pixels ($\ell\sim50$ scales), and compared to the CMB E-mode signal for $\tau=0.1$ (solid blue curve) and B-mode signal for $r=0.1$ (dashed blue curve). If an incorrect spectral index in synchrotron or thermal dust is assumed (e.g., by imposing a prior: tests F, G, H, and I), or a synchrotron curvature neglected (test C), the over- or under-subtraction of foregrounds at $\sim100-150$~GHz is significant compared to an $r=0.1$ signal.}
\label{fig:freq_depend}
\end{figure*}

This incorrect recovery in regions having a low signal-to-noise ratio is demonstrated in the left panels of Fig.~\ref{fig:betas_maps} for the synchrotron Q-Stokes component. Away from the Galactic plane, the index is estimated to be roughly $-2.5\pm0.5$. We also show in Fig.~\ref{fig:freq_depend} the frequency dependence of the components, rms averaged over the masked sky in $3.6^\circ$ pixels, and compared to the CMB signal in both E-modes and B-modes for $r=0.1$. Assuming that the synchrotron pivot is fixed at 30~GHz, an index that is too shallow by $\beta_s \sim 0.5$ overestimates the synchrotron power by of order $0.1~\mu$K in antenna temperature at the foreground minimum of 100~GHz. This is significant compared to the $r=0.1$ B-mode signal, so a bias is expected. 
Similarly for dust, with a pivot at 353~GHz, a dust emissivity index too steep by $\beta_d \sim 0.5$ would underestimate the dust at 100~GHz by up to $\sim 0.1~ \mu$K in antenna temperature; significant compared to the $r=0.1$ signal.

This specific case where the prior is systematically different to the input by up to 1$\sigma$ everywhere on the sky is a pessimistic scenario, but not implausible. To avoid the risk of bias, one must therefore take care in how the foreground model is parameterized. In the Bayesian framework, our chosen model has too many free parameters, given the low signal-to-noise ratio, so the result is being driven by the prior. To mitigate this, there are several ways of increasing the signal-to-noise ratio in the indices: including ancillary data from complementary experiments like {\it WMAP} and {\it C-BASS} \citep{King2010}, assuming common temperature and polarization spectral indices, using larger pixels to define the indices, or defining spectral indices in harmonic space to allow spatial coherence.  

\begin{figure}
  \centering
   \includegraphics[width=0.48\textwidth]{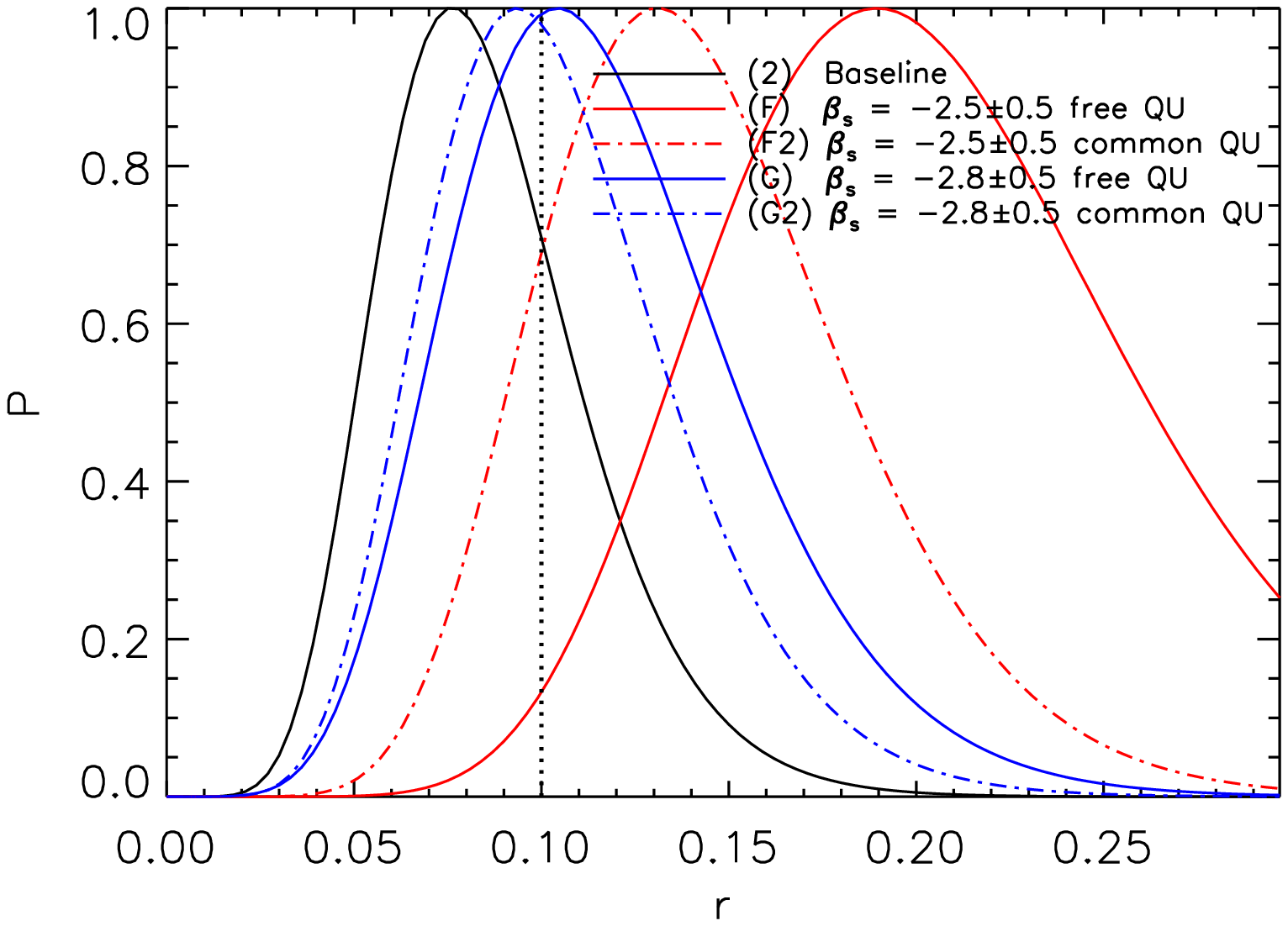} 
  \includegraphics[width=0.48\textwidth]{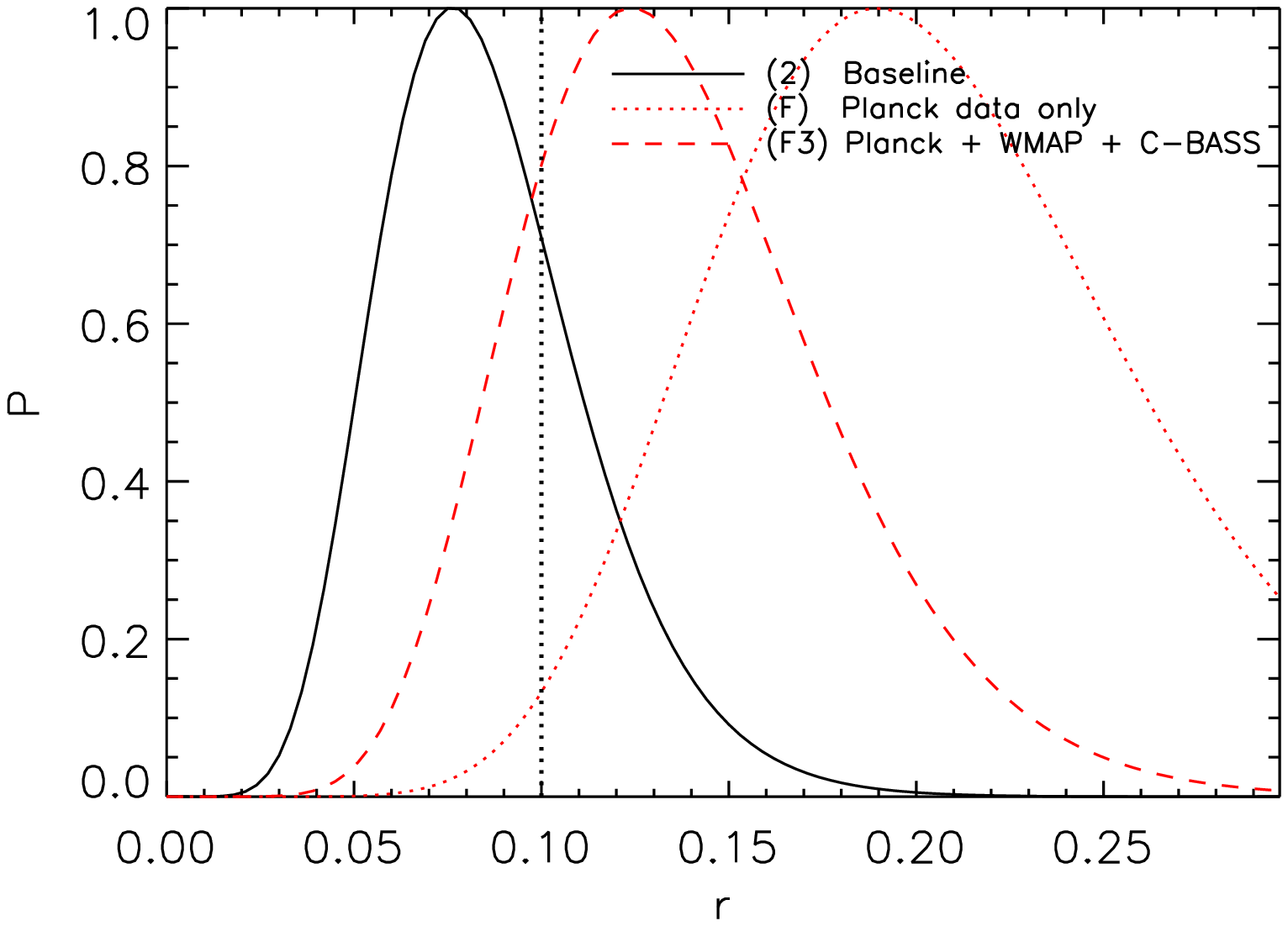}
   \caption{Recovered distributions for input $r=0.1$ for the baseline simulation with mean synchrotron index input $\beta_s=-3$, and Gaussian priors $-2.5\pm0.5$ or $-2.8\pm0.5$ (test F and G). The prior-dependent biases are reduced when the signal-to-noise is increased by assigning Q and U common indices (top, F2), or adding low-frequency data from {\it WMAP} or C-BASS (bottom, F3). }
\label{fig:additional_data}
\end{figure}

We consider two of these possible improvements. Each three-degree pixel can have a distinct spectral index for I, Q, and U. The first natural improvement is to fix the Q and U spectral indices to be common in each pixel, $\beta^s_Q=\beta^s_U$. Physically this is reasonable; the polarized signal comes from the same region of the Galaxy for both Q and U-type, and can be expected to have the same frequency dependence, consistent with observations \citep{Kogut07,Dunkley-WMAP,Gold09}. We repeat Tests F and G with this condition (Tests F2 and G2), and show the recovered index map in Fig.~\ref{fig:betas_maps}, with the likelihoods for $r$ in  Fig.~\ref{fig:additional_data}. The index map now has a higher signal-to-noise ratio, and the bias on $r$ reduced from more than $2\sigma$ to $1\sigma$ (for a prior of $\beta_s = -2.5\pm 0.5$). Fixing the temperature and polarization indices to be common is less physically motivated so we do not consider this here; depolarization effects could lead to different regions of the Galaxy contributing to the integrated polarization signal.

The signal-to-noise ratio can also be improved by adding ancillary data that better traces the foregrounds. Since the synchrotron signal dominates at lower frequencies, additional data at the low frequency range will increase the synchrotron signal-to-noise ratio.  We repeat Test F again (F3), adding simulated data from the {\it WMAP} 23~GHz K-Band channel, and projected {\it C-BASS} data at 5 GHz, to the simulated {\it Planck} data from 30-353 GHz. Figure \ref{fig:betas_maps} shows the significantly improved estimate of the synchrotron index in this case, which translates into a reduction in bias on $r$ from $2\sigma$ to $1\sigma$  for a prior of $\beta_s = -2.5 \pm 0.5$.  With the low frequency data, the indices are better constrained by the data.

A final obvious way to reduce the model freedom is to allow less spatial variation in the indices. In the limit of no spatial variation, this reduces to template cleaning \citep{Page07,Kogut07,EGP09}, with one spectral index over the whole sky. However, a concern with these methods is that they may not capture realistic spatial variation. The optimal balance is likely in between, requiring 
fewer than $\sim$3000 parameters to describe the spatially varying frequency dependence. Such an approach has been considered for polarization analysis in e.g., \citet{Dunkley-WMAP}, where 48 synchrotron spectral index parameters were used for {\it WMAP} component separation. In making this choice with real data, it will be important to test that results do not depend on the prior placed on frequency dependence. If so, the number of parameters should be reduced, or external data included where available.

\subsection{Effect of over-simplified model}
\label{subsec:oversimple}

In Section \ref{sec:wrong_model} we found that over-simplifying the frequency dependence of the two components can also lead to a bias in recovered parameters. Modeling the synchrotron as a power law everywhere on the sky, when it actually has a spectral curvature of $C=-0.3$, results in a $\sim 0.03$ bias high in $r$. As in Sec \ref{subsec:priors}, this can be understood as an overestimation of synchrotron at the 100~GHz range by up to $\sim 0.05 \mu$K in antenna temperature, illustrated in Fig.~\ref{fig:freq_depend}. Since some steepening is expected from synchrotron cooling, a strategy to prevent this bias would be to additionally marginalize over a curvature parameter. If the estimated CMB power does not change significantly with its inclusion, and the curvature is consistent with zero, this would justify neglecting the additional complexity. 

While we have examined only the case for synchrotron having a negative spectral curvature, there is some evidence to suggest that the spectral curvature could be positive \citep[e.g.,][]{Dickinson09,deOliveira-Costa08,Kogut07}.  This is not unexpected since multiple spectral components can give a flattening of the effective synchrotron index. With real data, a positive curvature as large as $0.3$ could be realistically considered.

At the high frequency end, thermal dust emission is typically modelled as a modified black-body, characterized by an emissivity and temperature, with $ I(\nu) \propto \nu^\beta B_\nu(T)$, and similarly for Q and U. This corresponds to our `one-component' dust model. A more complicated model has a sum of two or more components with different temperatures. In Sec \ref{sec:wrong_model} we found that modelling a two-component dust model as a one-component dust model has only a small effect on the estimated CMB signal. This reflects that the sum of two modified black-bodies, one sub-dominant, scales with frequency similarly to a single black-body. 

A larger bias was found for a modified black-body modelled as a power law. In this case we find a 1$\sigma$ shift in recovered $r$, with the power-law model typically over-subtracting dust. The effect is similar to neglecting synchrotron curvature. While it is unlikely in practice that the dust would be modelled as a pure power-law, it is possible that one could make the wrong choice for the dust temperature. In these tests we fixed the temperature to the input value that was common over the whole sky, and varied just the emissivity in each pixel. To check for a possible bias with real data, one would ideally additionally fit for the dust temperature.  Another approach to determine the dust temperature would be to use the temperature data, including the higher-frequency unpolarized channels of {\it Planck} (545 and 857~GHz), and IRAS/DIRBE data up to $\sim 3000$~GHz.  The dust temperature could then be assumed to be common for the polarization data.

\subsection{Effect of neglected components}

\begin{figure}
     \includegraphics[width=0.5\textwidth,angle=0]{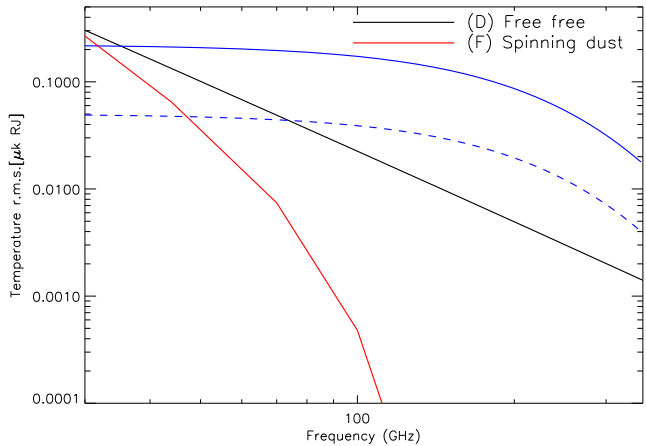}
      \caption{Frequency scaling of the 1\% polarized free-free and spinning dust foregrounds included in Tests D and E, rms averaged over the unmasked sky in $3.7^\circ$ pixels. At $\nu \gtsim 100$~GHz they are over an order of magnitude lower than the CMB E-mode signal for $\tau=0.1$ (solid blue curve), and below the B-mode signal for $r=0.1$ (dashed blue curve).}
\label{fig:freq_depend_ff_sd}
\end{figure}

We find that neglecting sub-dominant polarized free-free and spinning dust components has a negligible effect on the results. This can be understood from Fig.~\ref{fig:freq_depend_ff_sd}. The simulations include a 1\% polarized signal, with the rms signal of each component, averaged outside the Galactic mask, shown to be sub-dominant to an $r=0.1$ signal in the range $\nu>100$~GHz. The true polarization of these components is unknown, but is not expected to exceed this level. Observations of the Ophiuchi and Perseus cloud limit the polarization of spinning dust to be less than 2\% at 20-30~GHz \citep{dickinson11}, and WMAP observations limit it to less than 1\% over the whole sky. These levels are consistent with the spinning dust model by \citet{DL99}. For this mask, spinning dust polarization has a slightly larger effect on $r$ than free-free polarization. The spinning dust component is currently the most uncertain, so will be worth re-visiting with real data.

There are fewer observational constraints on the polarization of free-free emission. However, it should be intrinsically unpolarized because the scattering directions are random. Secondary polarization can be generated at the edges of bright free-free features from Thomson scattering \citep{RL79,Keating98}, but leading to less than $ 1$\% polarization at high Galactic latitudes. We have not therefore considered larger polarization levels.  We have also not considered more exotic components, such as a polarized `Haze' \citep{dobler/finkbeiner:2007}, or magnetic dust models \citep{DL99}. 

\section{Conclusions}
\label{sec:conclude}

Extracting robust estimates for the tensor-to-scalar ratio rely on modelling and subtracting polarized foregrounds. Since the polarized CMB signal is many times smaller than the foreground emission, the need to get this right is particularly acute. Many methods have been considered and implemented for foreground removal, but given the lack of data, the simulations are usually simple in form.

In this paper we have begun to quantify the impact on estimates of $r$ of incorrect foregound modelling. The tests were aimed at a detection of a signal with $r=0.1$, but the goal of future missions is to reach $r=0.01$ or lower, so we also consider an $r=0$ model. We conclude that neglecting a non-power-law frequency dependence of foregrounds may have a non-negligible effect on $r$; whereas neglecting a small free-free or spinning dust component is likely not to. 
We found that over-parameterizing the spectral indices had significant consequences; in the limit of a low signal-to-noise ratio the result can be highly prior-dependent.

We discussed methods of mitigating possible bias, through model comparison as more complexity is added to the foreground model, and through increasing the signal-to-noise ratio on spectral parameters by reducing their number and using ancillary data. We did not cover all scenarios of mismatch, but the approach of checking the goodness-of-fit through model comparison, and checking for a dependence of results on priors should be generally applicable. We did not explore the effects of different masks although this will be important to investigate with data \citep[see e.g.,][]{Dickinson09}.  Data from {\it Planck} and ground-based and balloon experiments will further elucidate the nature of the polarized foregounds and allow their modelling to be refined.  For full-sky data from future ultra-high sensitivity experiments such as {\it CMBpol} \citep{Bock09}, {\it COrE} \citep{Core11}, and {\it LiteBird} \citep{Litebird}, the effects studied here will be more important as we push towards $r=10^{-2} - 10^{-3}$ levels.

\medskip

We acknowledge the use of the Planck Sky Model, developed by the Component Separation Working Group (WG2) of the {\it Planck} Collaboration. We thank Aurelien Fraisse, Steven Gratton, and David Spergel for useful discussions. This work was performed using the Darwin Supercomputer of the University of Cambridge High Performance Computing Service (http://www.hpc.cam.ac.uk/), provided by Dell Inc. using Strategic Research Infrastructure Funding from the Higher Education Funding Council for England.  JD acknowledges support from ERC grant FPCMB-259505. CD acknowledges an STFC Advanced Fellowship and ERC grant under the FP7.

\end{document}